\documentclass[12pt]{article}
 \pdfoutput=1

\usepackage{amsmath,amssymb,amsfonts}
\usepackage{graphicx}
\usepackage{color}
\usepackage{amsmath}
\usepackage{amsfonts}
\usepackage{amssymb}
\usepackage{float}
\usepackage{cancel}
\usepackage{hyperref}
\usepackage{multirow}

\numberwithin{equation}{section}

\newcommand{\beq}{\begin{eqnarray}}
\newcommand{\eeq}{\end{eqnarray}}

\newcommand{\centeron}[2]{{\setbox0=\hbox{#1}\setbox1=\hbox{#2}\ifdim
\wd1>\wd0\kern.5\wd1\kern-.5\wd0\fi \copy0
\kern-.5\wd0\kern-.5\wd1\copy1\ifdim\wd0>\wd1
                                  \kern.5\wd0\kern-.5\wd1\fi}}
\newcommand{\ltap}{\>\centeron{\raise.35ex\hbox{$<$}}
                          {\lower.65ex\hbox{$\sim$}}\>}
\newcommand{\gtap}{\>\centeron{\raise.35ex\hbox{$>$}}
                          {\lower.65ex\hbox{$\sim$}}\>}

\newcommand\ZZ{\hbox{\zfont Z\kern-.4emZ}}
\font\zfont = cmss10 

\newcommand{\cref}[1]{Chapter \ref{c.#1}}

\newcommand{\met}{\mbox{${\rm \not\! E}_{\rm T}$}}

\def\beq{\begin{equation}}
\def\eeq{\end{equation}}
\newcommand{\ba}{\begin{array}}
\newcommand{\ea}{\end{array}}
\newcommand{\bea}{\begin{eqnarray}}
\newcommand{\eea}{\end{eqnarray} }
\newcommand{\bal}{\begin{align}}
\newcommand{\eal}{\end{align}}
\def\bi{\begin{itemize}}
\def\ei{\end{itemize}}
\def\ben{\begin{enumerate}}
\def\een{\end{enumerate}}
\def\beq{\begin{equation}}
\def\eeq{\end{equation}}
\def\bc{\begin{center}}
\def\ec{\end{center}}
\def\bt{\begin{table}}
\def\et{\end{table}}
\def\btb{\begin{tabular}}
\def\etb{\end{tabular}}

\def\mass2{mass${}^2$}

\begin{document}
\begin{titlepage}

\vskip.5cm
\begin{center}
{\huge \bf Slepton co-NLSPs at the Tevatron
 \vspace{.2cm}}

\vskip.1cm
\end{center}
\vskip0.2cm

\begin{center}
{\bf Joshua T.~Ruderman$^a$ and David Shih$^{b}$}
\end{center}
\vskip 8pt

\begin{center}
{\it $^a$Department of Physics\\
Princeton University, Princeton, NJ 08544}\\
\vspace*{0.1cm}
{\it $^b$ Department of Physics and Astronomy \\
Rutgers University, Piscataway, NJ 08854}\\
\vspace*{0.1cm}

\end{center}

\vglue 0.3truecm

\begin{abstract}
\vskip 5pt \noindent Ê

We study the Tevatron signatures of promptly-decaying slepton co-NLSPs in the context of General Gauge Mediation (GGM).  The signatures consist of trileptons plus $\met$ and same-sign dileptons plus $\met$.  Focusing first on electroweak production, where the Tevatron has an advantage over the early LHC, we establish four simple benchmark scenarios within the parameter space of GGM which qualitatively capture all the relevant phenomenology.    We derive limits on these benchmarks from existing searches, estimate the discovery potential with 10~fb$^{-1}$, and discuss ways in which these searches can be optimized for slepton co-NLSPs. We also analyze the Tevatron constraints on a scenario with light gluinos that could be discovered at the early LHC. Overall, we find that the Tevatron still has excellent reach for the discovery of SUSY in multilepton final states. Finally, we comment on the possible interpretation of a mild ``excess" in the CDF same-sign dilepton search in terms of slepton co-NLSPs.

\end{abstract}

\end{titlepage}

\newpage

\renewcommand{\thefootnote}{(\arabic{footnote})}

\tableofcontents

\section{Introduction}\label{sec:intro}
\setcounter{equation}{0} \setcounter{footnote}{0}

\subsection{Motivation}

Signatures involving several high $p_T$ leptons and missing energy are generally considered to be among the most promising channels in which to search for new physics at a hadron collider. Experimentally, these signatures are very clean. Leptons Ê(by which we mean electrons and muons, following standard experimental parlance) Êare straightforward to identify and difficult to fake. Moreover, the production rates for processes leading to several leptons plus missing energy are naturally small in the Standard Model, so backgrounds for new physics are low in these final states.

On the theoretical side, these signatures are also very well motivated. New physics scenarios
commonly give rise to multiple leptons and missing energy, especially in supersymmetric models. (For an overview of common supersymmetric models and their phenomenology, see~\cite{Martin:1997ns}.)
Perhaps the best known example occurs within the ``minimal supergravity" (mSUGRA) framework, in which the vast number of soft parameters of the MSSM are simplified down to just four continuous parameters. 
In this model, the lightest supersymmetric particle (LSP) is a bino-like neutralino, and it is absolutely stable. Pair producing heavier neutralinos and charginos and decaying them down to the LSP leads to final states with multiple leptons and missing energy.

In this paper we study a scenario that is at least as well motivated theoretically, although perhaps less known to experimentalists, in which multilepton plus $\met$ signatures also automatically arise. This is gauge mediated supersymmetry breaking (GMSB) with slepton co-NLSPs. (For a review of the phenomenology of gauge mediation and original references, see~\cite{Giudice:1998bp}.) 

\begin{figure}[t!]
\vbox{
\begin{center} \includegraphics[scale=0.6]{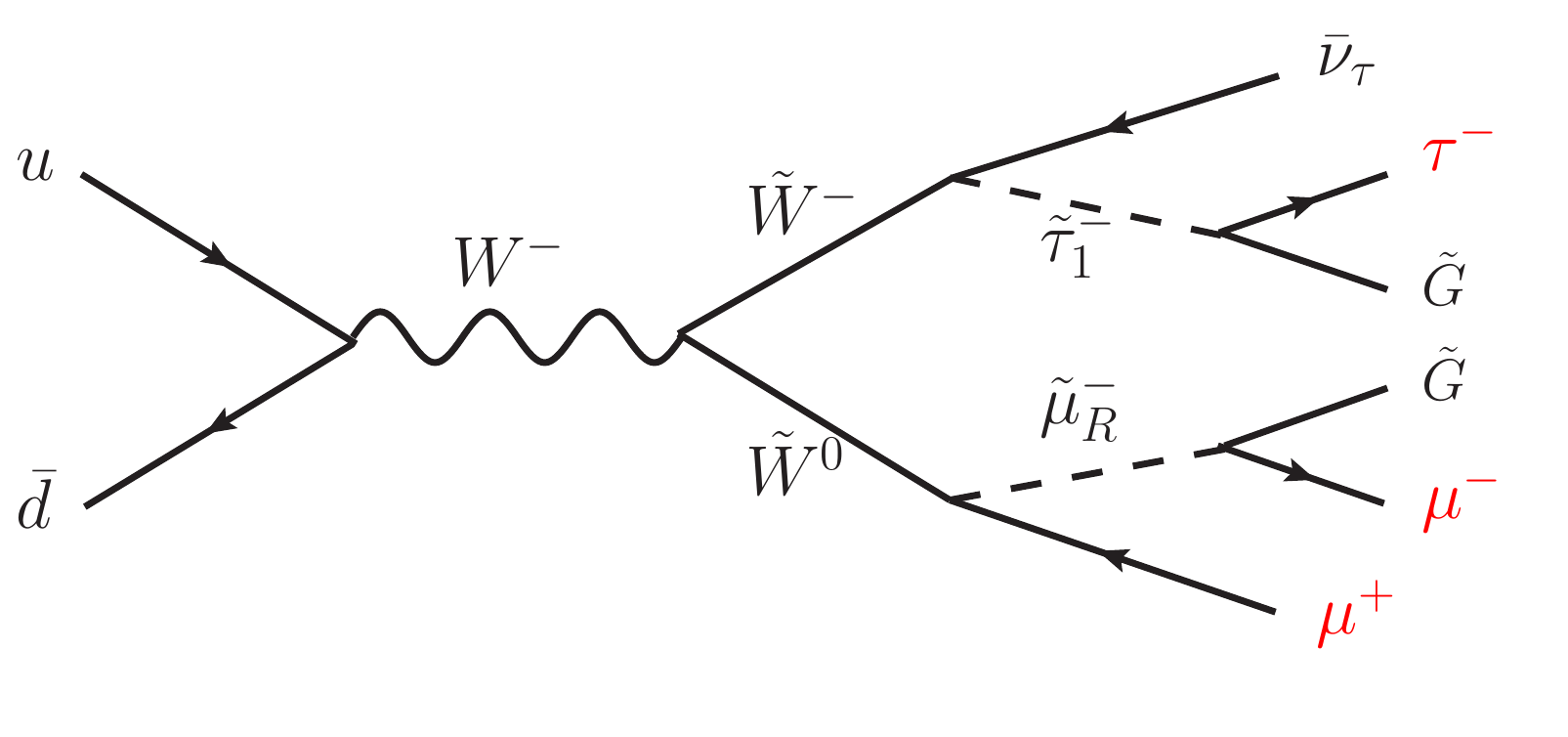} \end{center} 
\vspace{-0.7cm}}
\caption{An example slepton co-NLSP process leading to a trilepton final state. ÊHere, a charged and neutral wino are produced, and they decay through a right-handed smuon and stau, to produce a final state with $\mu^- \mu^+ \tau^- + \met$.} 
\label{fig:WinoDiagram}
\end{figure}

Recall that in gauge mediation, the LSP is a nearly-massless gravitino. The lightest MSSM sparticle is then the next-to-LSP (NLSP), and it always decays in a universal way to the gravitino plus its SM superpartner. Since this decay rate is heavily suppressed by the SUSY-breaking scale, all heavier sparticles decay first down to the NLSP before decaying to the gravitino. Therefore, the nature of the NLSP is the most important aspect of the GMSB spectrum for collider signatures. If the three flavors of right-handed sleptons all dominantly decay to gravitinos, and are thus co-NLSPs, then all MSSM events will contain at least two high $p_T$ leptons or taus, plus missing energy from the gravitinos.  Additional leptons can arise from cascade decays of heavier SUSY states, so that it is common to have three or more leptons plus taus in the final state (for an example see figure~\ref{fig:WinoDiagram}).  We focus on the situation where the sleptons decay promptly, which means gravitino masses in the sub-keV range. It is also interesting to consider delayed decays, but these signatures are entirely different and will be considered elsewhere. 

The importance of multilepton plus $\met$ signatures for the slepton (co-)NLSP scenario was first pointed out in a number of the original phenomenological studies of gauge mediation, including Refs.~\cite{Dimopoulos:1996vz,Dimopoulos:1996fj,Dimopoulos:1996yq,Ambrosanio:1997rv, Cheung:1997tg,Baer:1999tx,Qian:1998zs}, and by the Tevatron Run II SUSY working group~\cite{Culbertson:2000am}. ÊIn this paper, we revisit these signatures. This is motivated by two reasons.

First, the previous phenomenological work was focused on Minimal Gauge Mediation (MGM), a simplified model of GMSB in which the vast complexity of the MSSM is again reduced down to a handful of parameters. Recently in \cite{Meade:2008wd,Buican:2008ws}, a general framework of gauge mediation was formulated, and the full parameter space was shown to be much larger than that of MGM\@. 
While MGM enforces several large hierarchies among the soft masses (for example the left-handed sleptons are about twice as heavy as the right handed sleptons and the colored superparticles are significantly heavier than the uncolored ones), General Gauge Mediation (GGM) requires no such hierarchies, allowing new combinations of states to play important roles in collider signals. ÊThus it is interesting to revisit the collider signatures of GMSB within the full parameter space of GGM, in a model-independent way. 

Second, we now have the opportunity to confront the GMSB signals with the large existing dataset collected by the Tevatron. ÊUnfortunately, 
there was not much experimental follow-up to the early phenomenological work on GMSB with slepton co-NLSPs.
While these models were searched for systematically at LEP (we review the limits below), Tevatron searches optimized for GMSB have not been conducted in the multilepton plus $\met$ final states. However, related searches motivated by other scenarios (such as mSUGRA) have been conducted for same-sign dileptons \cite{D0Dilep, Abulencia:2007rd} and trileptons \cite{CDFTrilep3p2, Aaltonen:2008pv,Abazov:2009zi}. Thus it is interesting to examine the constraints these searches place on GMSB with slepton co-NLSPs, and to understand how to optimize their discovery potential. 

We emphasize that the Tevatron remains very relevant despite the advent of the LHC\@.  With 10~fb$^{-1}$, the Tevatron can produce more uncolored SUSY states than the LHC with $\lesssim 1~\mathrm{fb}^{-1}$ at 7 TeV, while many of the backgrounds are larger at the LHC\@. ÊThus it will remain the best place to search for the production of light uncolored states in the near term~\cite{Campbell:2010ff}. On the other hand, the LHC will very soon have the advantage for producing new light colored states. Therefore the two programs will remain complementary for the next several years.

\subsection{Summary of the main results}

Here we attempt to give a somewhat self-contained summary of our approach and our results, for the benefit of the more casual reader. Our philosophy and approach, motivated by GGM, is very much a continuation of the program begun in \cite{Meade:2009qv,Meade:2010ji}, where neutralino NLSPs were thoroughly studied in a general, model-independent way. (For other related references on collider signatures of GGM, see \cite{Carpenter:2008he, Rajaraman:2009ga, Abel:2009ve, Katz:2009qx, Katz:2010xg, Abel:2010vb}.) Ê

Even after restricting the NLSPs to sleptons, there are many possible spectra within GGM, and the parameter space is too large to fully explore. In order to reduce the complexity, our strategy is to choose spectra with as few light states as possible which play a role in the signal of interest, with other states decoupled. To that end, we focus on production of electroweak SUSY states -- neutralinos, charginos and sleptons -- and assume that the colored sparticles are too heavy to be produced. This is a natural assumption, given the relative strengths and weaknesses of the Tevatron vs.\ LHC described above. 

With the remaining electroweak SUSY states, we identify (in section \ref{sec:benchmarks}) four simple spectral types, classified primarily by the type of sparticle being produced. Specializing to gauge eigenstates, these correspond to: wino production, higgsino production, and left-handed slepton production with and without an intermediate bino. For each spectral type, we identify a two dimensional parameter space, spanned by the production mass and the NLSP mass, which together control the production cross-section and signal kinematics.  We choose parameters such that electrons, muons, and taus are produced democratically.

These simplified parameter spaces are intended to serve as benchmark scenarios which contain within them all the relevant Tevatron phenomenology of GMSB with slepton co-NLSPs. ÊThey are model-independent, in that they are defined using weak-scale soft parameters, and they are not subject to any arbitrary constraints or relations that exist in e.g.\ mSUGRA or MGM\@. Nevertheless we can be assured that our benchmark scenarios do correspond to physical models, since they conform to the parameter space of GGM \cite{Buican:2008ws}. Using these benchmark scenarios, we can investigate the phenomenological possibilities of slepton co-NLSPs. They can be used to quantify the strengths and weaknesses of existing experimental searches, and they can guide in the planning of future searches. 

To illustrate this, we use simulations to estimate the current limits and future (i.e.\ 10~fb$^{-1}$) discovery reach for each parameter space from the existing Tevatron searches in the trileptons$+\met$ and SS dileptons$+\met$ final states.\footnote{We use Pythia6.4~\cite{Sjostrand:2006za} to generate events, PGS4~\cite{PGS} for detector simulation, Suspect~\cite{Djouadi:2002ze} to calculate supersymmetric spectra, and a combination of Pythia and private code to calculate decay tables. For more details on our simulations, see appendix~\ref{app:calibratePGS}.}  Our main results are summarized in figs.\ \ref{fig:limits} and \ref{fig:discover} of section \ref{sec:results}. Apart from regions of parameter space where decays are kinematically squeezed, our results roughly translate to limits on the mass of the sparticle being produced:
\begin{eqnarray*}
&& m_{\tilde W} \gtrsim 190\,\,{\rm GeV},\quad m_{\tilde H} \gtrsim 190\,\,{\rm GeV},\quad m_{\tilde\ell_L}\gtrsim 155-170\,\,{\rm GeV} \qquad ({\rm current}\,\,{\rm limits}) \\
&& m_{\tilde W} \gtrsim 220\,\,{\rm GeV},\quad m_{\tilde H} \gtrsim 250\,\,{\rm GeV},\quad m_{\tilde\ell_L}\gtrsim 230-240\,\,{\rm GeV} \qquad (10\,\,{\rm fb}^{-1}\,\,{\rm reach})
\end{eqnarray*}
We find that the trilepton searches set the best current limits, owing to their larger dataset (3.2 fb$^{-1}$ vs.\ 1 fb$^{-1}$). However, we find that the CDF SS dilepton search has the best projected reach, owing to its more inclusive nature.

Also in section \ref{sec:results}, we show that the Tevatron still has $3\sigma$ discovery potential for a wide range of superpartner masses consistent with the current limits, and $5\sigma$ discovery potential for a narrower range of allowed masses, especially if the searches are optimized for the kinematics of the slepton co-NLSP signal. Ê

In section \ref{sec:future}, we go beyond our four benchmark scenarios to discuss various promising directions for future research. 
First, we consider the degradation in experimental sensitivity for models that preferentially produce taus instead of electrons and muons, such as models with stau NLSP\@.  Overall, we find that the limit and reach for such models is currently quite poor, given the higher backgrounds and lower efficiencies. The trilepton searches do end up being more sensitive than the SS dilepton searches, because they include one-prong hadronic $\tau$ decays in their list of final states.  Second, as preliminary to studying colored production at the LHC, we present the Tevatron limits and reach for a slepton co-NLSP scenario with gluino production. In a forthcoming paper~\cite{JoshDavidLHC}, we will study the phenomenology of this and other GGM scenarios in much more detail.

Section~\ref{sec:discussion} contains our concluding discussion.  Here, we suggest ways to optimize the Tevatron multilepton searches for GMSB, guided by our results in the previous sections. 

\begin{itemize}

\itemÊThese searches have been biased in various ways to accommodate the kinematics of the mSUGRA signal. ÊAs we will see, GMSB with slepton co-NLSP produces leptons with larger $p_T$'s and larger missing energy than mSUGRA\@ (a comparison appears in appendix~\ref{sec:GGMvSugra}).Ê Thus the cuts on these quantities Êcan be easily tightened without losing much signal acceptance. This could also help improve the sensitivity in channels that include hadronic $\tau$ decays (e.g.\ the single track category of the trileptons search).

\item In GMSB it is quite common to have more than three leptons (up to 6-8 depending on the model!) in an event, coming from cascade decays of heavier states~\cite{Dimopoulos:1996fj,Cheung:1997tg,Baer:1999tx}. Thus it is important to be as inclusive as possible and not veto e.g.\ on additional leptons in the event. 

\item Similarly, one should avoid vetoing on leptons forming $Z$ bosons if at all possible, since with so many leptons in an event, it is quite common that two will accidentally reconstruct a $Z$. 

\item Finally, we point out that there are two major gaps in Tevatron coverage of slepton (co-)NLSPs. First, there are no existing Tevatron searches for $\ge 4$ leptons plus missing energy, which is known to be a promising channel for discovery of GMSB \cite{Baer:1999tx}. Second, searches for SS dileptons which include hadronic  tau decays  would greatly improve the coverage of tau-rich scenarios, despite the higher backgrounds.

\end{itemize}


Finally, we explore (also in section~\ref{sec:discussion}) the intriguing possibility that a hint of an excess in the 1~fb$^{-1}$ CDF same-sign dilepton search \cite{Abulencia:2007rd} is due to slepton co-NLSPs. The slight ``excess" (not statistically significant) occurred at high $\met$ and also at high lepton $p_T$. Specifically, CDF saw four events with $\met>80$~GeV and four events with lepton $p_T>80$~GeV where they expected $\sim 1$. While these events most likely came from insufficiently understood background tails, it is entertaining to consider their possible origin in our slepton co-NLSP scenarios. As discussed above, events with high $\met$ and high lepton $p_T$ are very characteristic of slepton co-NLSPs. Indeed, we find that if these excesses are real, slepton co-NLSPs can easily explain them consistent with current experimental limits, especially the slepton-production benchmark scenarios where the existing limits are much weaker. Moreover, if these excesses are real and are explained by slepton co-NLSPs, they will show up spectacularly in 10 fb$^{-1}$ as a 5$\sigma$ discovery.

Our appendices include technical details and special topics. Appendix~\ref{app:calibratePGS} contains details of our simulations.  In appendix~\ref{sec:MGM}, we apply our general results to the special case of MGM with slepton co-NLSPs. This is contained within just one of our benchmark scenarios (slepton production with intermediate bino), and we derive the corresponding limits on the MGM parameter space.  Finally, in appendix~\ref{sec:GGMvSugra} we compare the multilepton signals of GMSB to those of mSUGRA\@.

\section{Slepton co-NLSPs in GGM}\label{sec:taxonomy}

\subsection{GGM Parameter Space and Production Modes}

We begin by reviewing the parameter space of GGM~\cite{Meade:2008wd,Buican:2008ws}.  In GGM, the spectrum is determined by the five sfermion mass-squareds, $m_{\tilde Q}^2$, $m_{\tilde u}^2$, $m_{\tilde d}^2$, $m_{\tilde e_L}^2$, $m_{\tilde e_R}^2$, which are subject to two sum rules\footnote{The GGM sum rules (see Ref.~\cite{Meade:2008wd}) do not play an important role here.  This is because, for simplicity, we consider spectra with a small number of light superpartners that produce the signal of interest, with the other superparticles decoupled, $m \gtrsim$~TeV\@.  For the examples we consider, the decoupled superparticle masses can always be chosen such that the sum rules are satisfied.}; and three gaugino masses, $M_1$, $M_2$, $M_3$, not subject to any relations.  The sfermion mass-squareds are flavor degenerate at the messenger scale, but the first two generations become split from the third via RG evolution. (For a nice summary of the MSSM RGEs, see \cite{Martin:1997ns}.) The Higgs soft masses $m_{H_u}^2$ and $m_{H_d}^2$ are sensitive to corrections from additional sector(s) that e.g.\ address the $\mu$ problem (see \cite{Meade:2008wd, Komargodski:2008ax} for discussions of this in the context of GGM). So we treat these as free parameters, and trade them in practice for $\mu$ and $\tan\beta$ defined at the weak scale. Finally, the $A$-terms are zero at the messenger scale (and hence small at the weak scale), and the gravitino mass is $\ll m_{weak}$.

\begin{figure}[!t]
\begin{center}
 \includegraphics[scale=0.5]{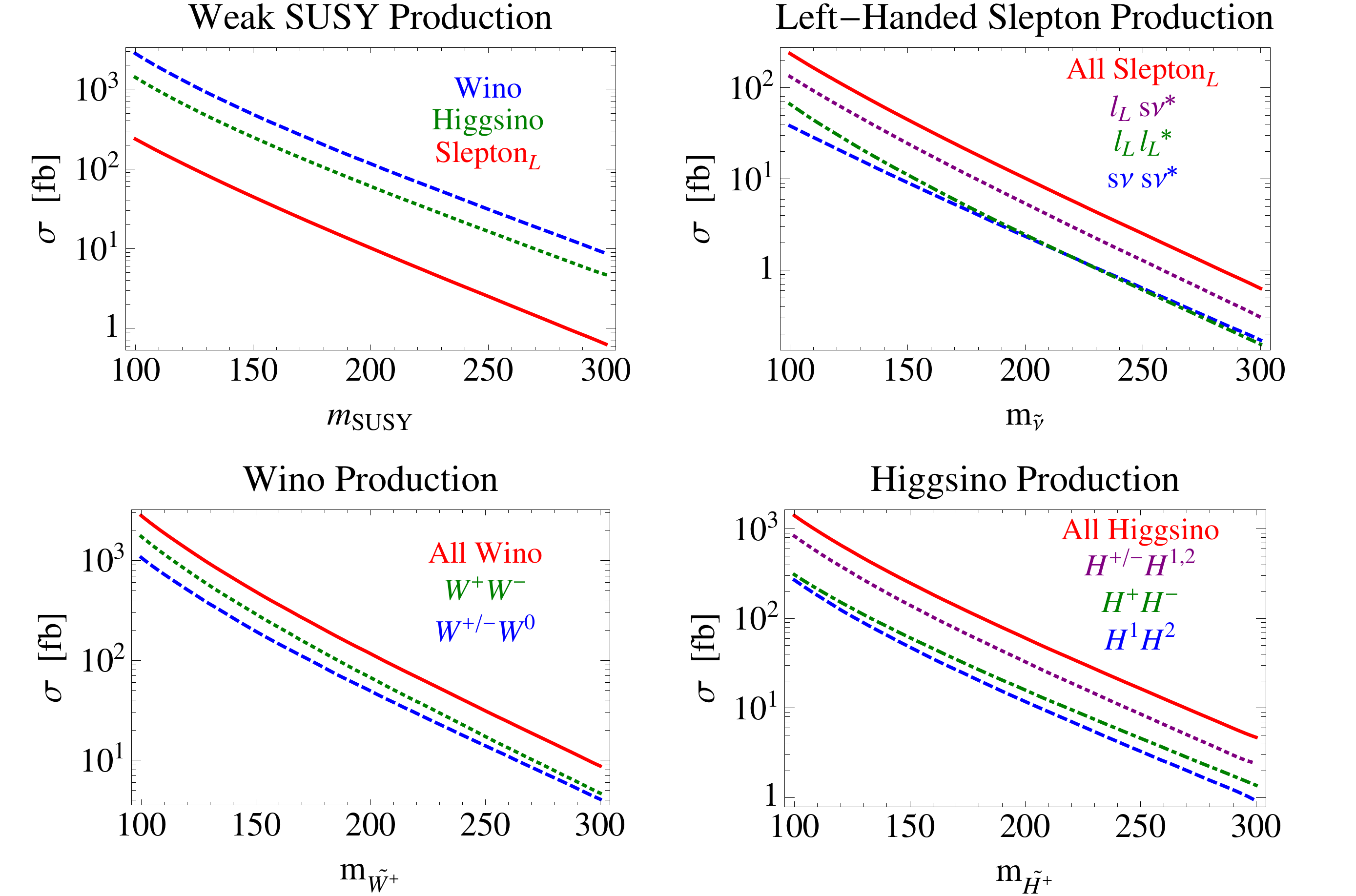}
\caption{\label{fig:prod} Tevatron production cross-sections (LO) for winos, higgsinos, and left-handed sleptons.  For all plots, we take $\tan \beta=2$.  For left-handed slepton production, we take flavor symmetric soft terms at the weak scale, and we fix $\mu=300$ and $m_{\tau_R}=100$ GeV\@.  These parameters enter the stau mixing.  For wino production, we fix $\mu=M_1=1$~TeV\@.  For higgsino production, we fix $M_1=M_2=1$~TeV\@.}
\end{center}
\end{figure}

As discussed in the introduction, since we are focusing on the Tevatron in this paper, we will assume electroweak sparticle production only, with colored sparticles decoupled, $m_{colored}\gtrsim 500$ GeV\@.\footnote{Near the end of the paper, we will also briefly consider the Tevatron limits and reach for colored sparticle production, with an eye towards discovery at the LHC\@.} So the relevant parameters are:
\begin{equation}
\label{eq:param}
 M_1, \quad M_2, \quad \mu, \quad \tan \beta, \quad m_{\tilde e_L}, \quad m_{\tilde e_R}, \quad m_{\tilde \tau_L}, \quad m_{\tilde\tau_R}
\end{equation}
Keeping in the spirit of our model-independent philosophy, we choose to specify these parameters at the weak scale and remain agnostic as to their UV (messenger scale) origin.

Potential SUSY production modes involve the neutralinos and charginos, or the left-handed sleptons (both charged sleptons and sneutrinos). We specialize for simplicity to gauge eigenstate -inos.
Plots of wino, higgsino, and slepton production cross-sections are shown in fig.~\ref{fig:prod}. Higgsinos and winos have similar cross-sections, with $\sigma \sim 100$~fb for $m_{\tilde H} \sim m_{\tilde W} \sim 200$~GeV\@.  The total slepton production cross-section is a factor of 5-10 lower at the same mass, e.g.\ it is $\sim10$~fb for 200 GeV sleptons. Not shown here is the cross-section for right-handed sleptons, which is further suppressed. As we will discuss in section \ref{subsec:leplimit}, $\tilde\ell_R^+\tilde\ell_R^-$ production  leads solely to OS dileptons plus $\met$, and the large SM backgrounds for this final state, combined with the smaller SUSY cross section, make this an unsuitable search channel at the Tevatron.

\subsection{Slepton co-NLSP Definition}

Now that we have defined the parameter space, we are ready to discuss what it means for a spectrum to have slepton co-NLSPs.  We specialize to spectra where the right-handed sleptons are at the bottom of the spectrum.\footnote{We do not consider spectra with left-handed sleptons at the bottom, because the electroweak $D$-term splits the slepton doublet, causing the sneutrinos to be the NLSP\@.  
The collider signatures of this scenario have been studied elsewhere~\cite{Katz:2009qx,Katz:2010xg}.}  Examples of such spectra are shown in fig.\ \ref{fig:spectra}; these will be discussed in more detail in section \ref{sec:benchmarks}.
Since  the right-handed slepton masses originate from a flavor universal boundary condition at the messenger scale, the right-handed selectron and smuon are always very nearly degenerate.  The lightest stau, $\tilde \tau_1$, is always the lightest state, because of RG running and left-right stau mixing. 

Whereas $\tilde \tau_1$ will always decay to a $\tau$ and gravitino, the other sleptons have two possible decay channels \cite{Dimopoulos:1996fj}.  For smaller $\delta m_{\tilde l_R} =  m_{\tilde e_R}-m_{\tilde \tau_1}$, the first two generation sleptons dominantly decay to a lepton and gravitino, and this defines the slepton co-NLSP regime.  
In a collider, every SUSY event contains at least two hard leptons or taus (depending on whether $\tilde l_R$ or $\tilde \tau_1$ is produced), and significant missing energy, carried by the gravitinos.  For larger $\delta m_{\tilde l_R}$, the first two generations dominantly decay three-body, to $\tilde \tau_1$ through an off-shell neutralino, $\tilde l_R^{\pm} \rightarrow l^{\pm} N_i^* \rightarrow l^{\pm} \tau^{\mp} \tilde \tau_1^{\pm}$.  Every SUSY cascade ends in $\tilde \tau_1$, and this defines the stau NLSP regime. 

Both the slepton co-NLSP and the stau NLSP scenarios are ``generic" in models of gauge mediation, i.e.\ no special fine tuning of UV parameters is required to realize either scenario. In general, the slepton co-NLSP regime applies for $\delta m_{\tilde l_R} \lesssim 5-10$ GeV, while the stau NLSP regime holds for larger mass splittings \cite{Ambrosanio:1997bq}. The mass splitting in turn is controlled primarily by $\tan\beta$ and the left-right mass splitting, as this sets the amount of left-right mixing for the third generation sleptons. For example, in MGM $\tan\beta\lesssim 10-15$ corresponds to $\delta m_{\tilde l_R} \lesssim 5-10$~GeV~\cite{Dimopoulos:1996yq}. Since we will focus on the slepton co-NLSP regime, we will henceforth take
\begin{eqnarray}
m_{\tilde e_R} = m_{\tilde \tau_R} \qquad m_{\tilde e_L} = m_{\tilde \tau_L} 
\end{eqnarray}
This simplifies the parameter space, without much loss of generality.

\subsection{Prequel: the LEP Limit}
\label{subsec:leplimit}

Before we identify benchmark slepton co-NLSP scenarios for the Tevatron, and the attendant limits for each, we take a moment to review the limit set by LEP2.  The LEP experiments looked systematically for gauge mediation with slepton co-NLSPs with either prompt or displaced decays.  The results are compiled and summarized by the LEP2 SUSY working group~\cite{LEP2}.  Slepton co-NLSPs with prompt decays were searched for in the acoplanar leptons final state (OS dileptons plus missing energy).  The strongest model-independent limit comes from direct $\tilde \mu_R \tilde \mu_R^*$ production, requiring $m_{\tilde \mu_R} > 96$~GeV at 95\% confidence level.  Right-handed stau production is less-constrained, $m_{\tilde \tau_1} > 87$~GeV, because taus have a lower experimental acceptance, and the limit on $\tilde e_R$ depends on the neutralino spectrum, because of possible interference between the $s$ and $t$-channels~\cite{Ambrosanio:1997rv}. Within GGM, the limit from acoplanar muons sets the strongest limit on the right-handed selectron mass, because $ m_{\tilde e_R}=m_{\tilde \mu_R} > 96$~GeV follows from the flavor-independent boundary condition, discussed above.

OS dilepton plus missing energy is not the most promising channel to look for slepton co-NLSPs at the Tevatron. 
This is because there are significant backgrounds due to leptonically decaying $W^-W^+$ and $t\bar t$, with $\sigma\times {\rm Br} \approx 0.6$~pb and $\sigma\times{\rm Br}\approx 0.4$~pb, respectively.  
As we will review in section \ref{sec:searches}, the backgrounds are much, much lower in final states with same-sign dileptons or trileptons. Thus, we will focus on these more promising final states in the rest of the paper. The trade-off here is that they are only populated through cascade decays of heavier SUSY states. Thus searches in these final states will necessarily be less inclusive than the LEP searches, and describing their constraints will require taming the zoo of possible GGM spectra that contain slepton co-NLSPs at the bottom. We will turn to this problem in the next section.

\section{Benchmark Scenarios for Slepton co-NLSPs}
\label{sec:benchmarks}

As discussed above, the Tevatron sets the strongest limits on slepton co-NLSPs through the same-sign dilepton and trilepton channels.  These final states are produced by SUSY cascades ending in right-handed sleptons, but the number of leptons in the final state, and their kinematics, are sensitive to the details of the SUSY spectrum.  This means that there are a very large number of possibilities in GGM, spanned by the 8-dimensional parameter space identified in equation~\ref{eq:param}.  We are unable to cover this full space, but we can attempt to cover all the qualitative phenomenological possibilities. To that end, we have identified four model-independent benchmark scenarios for slepton co-NLSPs at the Tevatron.  By studying them and optimizing experimental searches for them, one will (hopefully) ensure that no large qualitative gaps remain in the experimental coverage of the GGM parameter space. 

\begin{figure}[t!]
\vbox{
\begin{center}
\includegraphics[width=.7\textwidth]{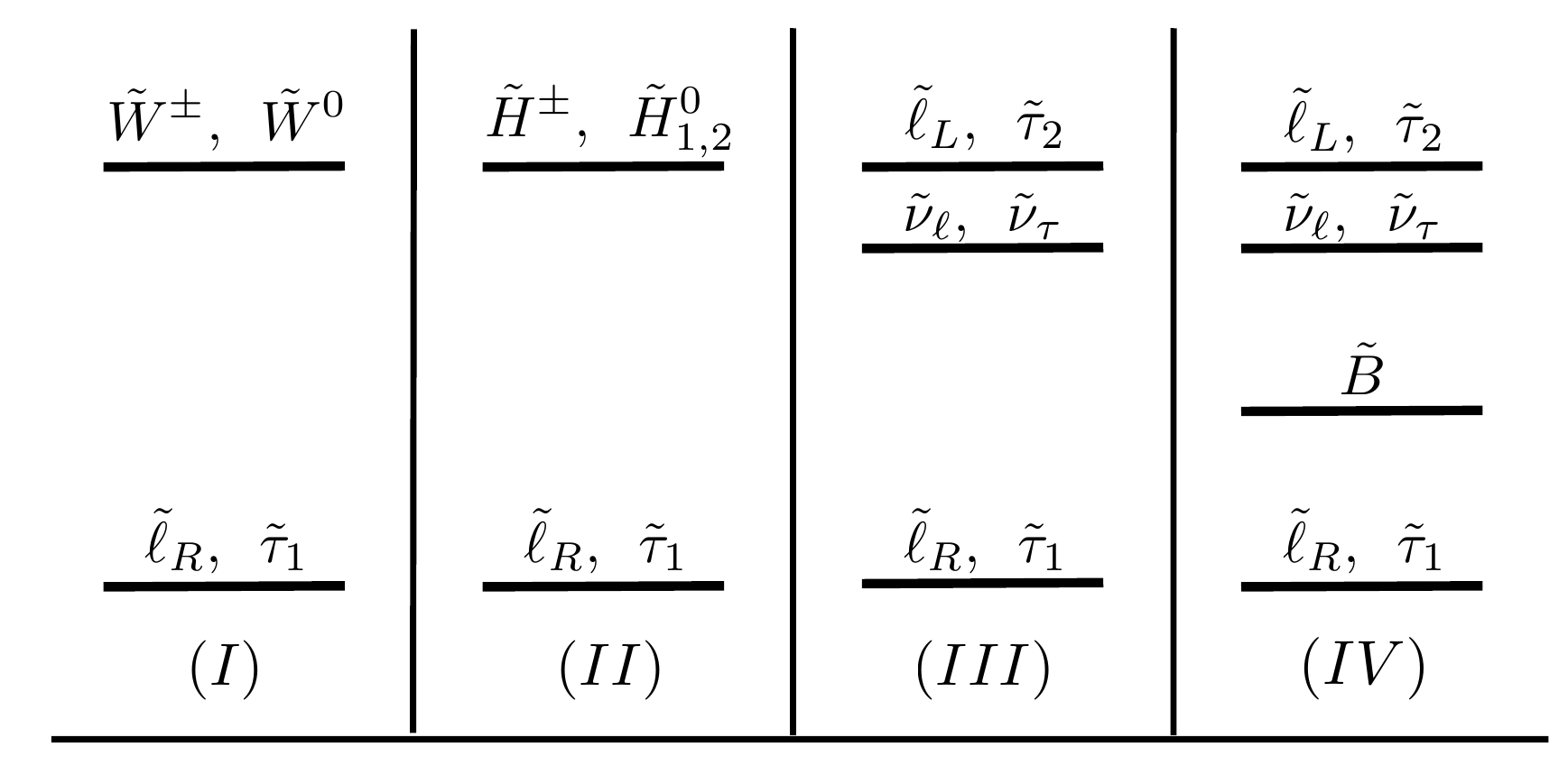} 
\end{center}
\vspace{-0.75cm}
}
\caption{The different slepton co-NLSP spectra that form our benchmarks.}\label{fig:spectra}
\end{figure}

Our simplifying approach is to consider spectra with one type of production mode, which then decays to right-handed sleptons.  The possibilities are shown in figure~\ref{fig:spectra}: wino production, higgsino production, and left handed slepton production with and without an intermediate bino.   For each benchmark scenario we choose parameters, shown in table~\ref{tab:bench}, that maximize the branching ratio to leptons $l=e,\mu$, which are more experimentally constrained than taus.  We stress that we have chosen the minimal spectra that produce same-sign dilepton and trilepton final states.  It is possible to combine our benchmarks with additional heavier states that produce more elaborate decay chains.  Our results correspond to conservative limits on such scenarios.

\begin{table}[!t]
\begin{center}
\begin{tabular}{|c|c|c|c|c|}
\hline
& \multirow{2}{*}{(I) Wino prod.} & \multirow{2}{*}{(II) Higgsino prod.} & \multirow{2}{*}{(III) Slepton prod.} & (IV) Slepton prod. + \\
& & &  & intermediate bino  \\
\hline
$m_{\tilde e_L}= m_{\tilde \tau_L}$ & 1000 
& 1000 
& vary
& vary
 \\ 
$m_{\tilde e_R} = m_{\tilde \tau_R}$ & vary 
& vary 
& vary
& vary
\\
$M_1$ & 300   
& 300  
& 300
& $(m_{\tilde e_L}+m_{\tilde e_R})/2$
\\
$M_2$ & vary 
& 1000
& 1000
& 1000
\\
$\mu$ & 1000 
& vary $(\mu<0)$ 
&
500
&
500
 \\
$\tan\beta$ & 2   
 & 2  
 & 2
 & 2 
 \\
\hline
\end{tabular}
\caption{Benchmark scenarios for slepton co-NLSPs with flavor democratic decays.  Masses are specified in GeV. We choose a low-scale for SUSY breaking, $\sqrt F = 50$~TeV, giving prompt decays to the gravitino. 
\label{tab:bench}}
\end{center}
\end{table}

\subsection{Wino}

The first benchmark we consider consists of winos that are heavier than right-handed sleptons.  Since there is no coupling between charginos and right-handed sleptons, the charged wino always decays to $\tilde\tau_1$,
\begin{equation}
W^\pm \to \tilde\tau_1^\pm + \nu_\tau
\end{equation}
Meanwhile, the neutral wino can decay down to all three generations of right-handed sleptons:
\begin{equation}
 W^0 \to \tilde e_R^\pm+e^\mp,\,\,\,\tilde\mu_R^\pm+\mu^\mp,\,\,\,\tilde\tau_1^\pm+\tau^\mp
 \end{equation}
The decays to the first two generations happen only through mixing with the bino. Decays down to $\tilde\tau_1$ can be mediated in the same way, through mixing with the higgsino, or through stau mixing. The latter two are proportional to the tau Yukawa coupling, so they are enhanced at large tan $\beta$. By varying this and other parameters, we can make the neutral wino branching fractions to the right-handed sleptons flavor democratic, or tau-dominated.  For the benchmark scenario, we choose parameters, shown in table~\ref{tab:bench}, that lead to democratic decays: low $\tan \beta$ and $M_1 \ll \mu$.  

We see that charged wino pair production, $\tilde W^+ \tilde W^-$, leads solely to an opposite-sign ditau final state, and not same-sign dileptons or trileptons.  These final states do result from the associated production of a neutral and charged wino, $\tilde W^0 \tilde W^\pm$. For this benchmark,  $2/3$ of $\tilde W^0 \tilde W^\pm$ events have two OSSF leptons and $\tau^\pm$, while $1/3$ of the events have three $\tau$'s.

One simple way to modify this benchmark would be to add an intermediate neutralino that is mostly bino, $m_{\tilde l_R} < M_1 < M_2$.  Then, the mostly-wino neutralino can decay to the mostly-bino neutralino plus an on or off-shell $Z$.  In most of the parameter space that will be of interest at the Tevatron, this decay is three-body and subdominant to the decays discussed above, and we therefore choose not to consider this situation separately. 

\subsection{Higgsino}

Now we consider a benchmark with higgsinos heavier than the right-handed sleptons.  Recall that a spectrum with mostly-pure higgsinos consists of two neutral higgsinos, $\tilde H_{1,2}$, and a charged higgsino, $\tilde H^\pm$, all nearly degenerate with one another. Similarly to the wino benchmark, the charged higgsino can only decay to $\tilde \tau_1$ plus a neutrino, whereas the two neutral winos, $\tilde H^0_{1,2}$, can decay to $\tilde e_R^\pm e^\mp$, $\tilde \mu_R^\pm \mu^\mp$, or $\tilde \tau_1^\pm \tau^\mp$.  
The main difference between higgsino and wino production is that higgsinos include a new production mode of two neutral higgsinos, $\tilde H^0_1 \tilde H^0_2$, which results in a 4-lepton/tau final state.  We choose parameters, shown in table~\ref{tab:bench}, that yield flavor democratic decays of the neutral higgsinos: low $\tan \beta$ and low $M_1$.  The production of a neutral plus charged higgsino results in the same final states as the wino benchmark. When two neutral higgsinos are produced, $4/9$ of the events have two OSSF lepton pairs, $4/9$ have one OSSF lepton pair and two $\tau$'s, and $1/9$ of the events have four taus.    Similarly to the discussion for the wino benchmark, we could augment the spectrum with an intermediate bino, but the SUSY cascades will bypass the bino for most of the parameter space of interest for the Tevatron, yielding signals similar to this benchmark.

\subsection{Left-Handed Slepton}

The third scenario we study consists of left-handed sleptons that are produced and decay to right-handed sleptons.  The production modes are $\tilde l_L^\pm \tilde \nu$, $\tilde \nu \tilde \nu^*$, and $\tilde l_L^- \tilde l_L^+$.  Recall that the left-handed sleptons and sneutrinos are split by the electroweak $D$-terms, with the former  always slightly heavier, 
\begin{equation}
m_{\tilde e_L}^2 - m_{\tilde \nu}^2 = \left|\cos 2 \beta \right| m_W^2
\end{equation}
We will now discuss how $\tilde l_L$ and $\tilde \nu$ decay. The decays are quite different for the first two generations and for the third generation, so we will discuss them separately.

The first two generation left-handed sleptons can three-body decay to the right-handed sleptons via an off-shell neutralino,
\begin{eqnarray}
\label{lLdectwo}
&& \tilde \ell_L^-  \quad\to\quad \tilde e_R^\pm + e^\mp + \ell^-,\quad \tilde \mu_R^\pm + \mu^\mp + \ell^-,\quad \tilde \tau_1^\pm + \tau^\mp + \ell^- 
 \end{eqnarray}
Depending on the -ino masses and $\tan\beta$, these decays can either be flavor democratic or tau-rich. Again, for our benchmark scenario, we choose parameters (shown in table~\ref{tab:bench}) such that these decays are flavor democratic. We also see from (\ref{lLdectwo}) that the decay of $\tilde\ell_L^-$ can lead to either a SS or OS lepton pair. The relative branching fraction depends on the ratio of the off-shell neutralino mass to the slepton mass, with the SS dominating as this ratio is increased~\cite{Ambrosanio:1997bq}.

The above decays typically dominate, except when the left-handed and right-handed sleptons are squeezed or when the neutralinos are too heavy.  In these cases, the left-handed sleptons prefer to decay to the sneutrino through an off-shell $W$,
\begin{equation}
\label{lLdecone}
 \tilde \ell_L^- \to \tilde\nu_{\ell}+W^{-*} 
 \end{equation}

The sneutrino decays are relatively simpler. For the first two generations, the only possibilities are again three body decays through off-shell neutralinos:
\begin{eqnarray}
\label{nuLdecone}
&&\tilde\nu_\ell  \quad \to\quad \tilde e_R^\pm + e^\mp + \nu_\ell ,\quad \tilde \mu_R^\pm + \mu^\mp + \nu_\ell,\quad \tilde \tau_1^\pm + \tau^\mp + \nu_\ell
\end{eqnarray}
As above, depending on the -ino masses and $\tan\beta$, these decays can either be flavor democratic or tau-rich, and our choice of parameters in table~\ref{tab:bench} leads to flavor democratic decays. 

Finally we come to the third generation left-handed sleptons. While three-body decays analogous to Eq.~\ref{lLdectwo} are also available for the left-handed stau, $\tilde \tau_2$, the stau can also decay through tau-mixing and an on or off-shell $Z$,
\begin{equation}
\label{lLdecthree}
\tilde\tau_2^- \to \tilde\tau_1^- + Z^{(*)}
\end{equation}
For the parameters of table~\ref{tab:bench}, this is the dominant decay mode until $\tilde \tau_1$ and $\tilde \tau_2$ become squeezed, and then Eq.~\ref{lLdecone} dominates. Also, for these benchmark parameters, the third-generation sneutrino prefers to decay through stau mixing via an on-shell or off-shell $W$,
\begin{equation}
\label{nuLdectwo}
\tilde\nu_\tau \to \tau_1^- +  W^{+(*)}
\end{equation}
Evidently,  the third generation sleptons dominantly produce OS tau pairs, as can be seen by inspecting equations~\ref{lLdecone}, \ref{lLdecthree}, and \ref{nuLdectwo}. (Additional leptons can come from the $W$ and $Z$'s, but then one must pay the branching fraction price, and plus these are generally softer.)   Due to the lower experimental sensitivity to taus, we find that the production of the third-generation sleptons can be ignored when setting limits,  relative to the production of the first two generations.

\subsection{Left-Handed Slepton with Intermediate Bino}
The left-handed slepton decays are significantly simpler if there is a bino between the left-handed and right-handed sleptons.  This situation deserves special consideration and constitutes our fourth and final benchmark (we pick parameters shown in table~\ref{tab:bench}).  Now, the left-handed slepton states, $\tilde l_L$ and $\tilde \nu$, all decay down to the bino plus their superpartner, and the bino decays flavor democratically and charge democratically to the right-handed sleptons.  This means that the three-body decays through off-shell $Z$ and $W$ (Eqs.~\ref{lLdecone}, \ref{lLdecthree}, and \ref{nuLdectwo}) are bypassed, resulting in more leptons in the final state and, as we will see, a stronger experimental limit.

\subsection{Benchmark Collider Signals}

We conclude this section by giving a brief characterization of the multilepton collider signals of the four benchmark models.  Table~\ref{tab:topologies} summarizes the above discussion, listing the relevant production modes and decay topologies for each benchmark. Here we are assuming that the spectra are unsqueezed, but other than this assumption, the topologies are independent of the NLSP and production masses. The final two columns show the percentage of events, weighted by the $\sigma \times\mathrm{Br}$ of each topology, that contain same-sign dileptons or trileptons, for $m_{e_R}\approx 150$ GeV, $m_{prod}\approx 200$ GeV. ($m_{prod}$ denotes the mass of the particles being produced: wino, higgsino or slepton.)
These leptons (again, $e$ or $\mu$) were required to have $|\eta|<1$ and $p_T>10$ GeV, and were simulated using the default PGS4 CDF simulation (so they include crude lepton ID cuts and parametrization of central calorimeter cracks)~\cite{PGS}.  

\begin{table}[!t]
\begin{center}
\begin{tabular}{|c|c|l|c|c|}
\hline
 Benchmark scenario & Prod.\ mode &  Topologies &  \% SS Dilepton  & $\ge 3$ Leptons \\
\hline
 Wino & $\chi_1^\pm\chi_1^0$ & $2\ell\,1\tau,\,\,0\ell\,3\tau $ & 8.8 &  5.3 \\
\hline
\multirow{2}{*}{Higgsino}  & $\chi_1^\pm\chi_{1,2}^0$   & $2\ell\,1\tau,\,\,0\ell\,3\tau $& \multirow{2}{*}{16.5} & \multirow{2}{*}{11.9}   \\
& $\chi_1^0\chi_2^0$ & $4\ell\,0\tau,\,\,2\ell\,2\tau,\,\,0\ell\,4\tau$ & & \\
\hline
\multirow{6}{*}{Slepton}  & $\tilde{\nu}_\ell\tilde{\nu}_\ell$   & $4\ell\,0\tau,\,\,2\ell\,2\tau,\,\,0\ell\,4\tau$ & \multirow{6}{*}{38.7} & \multirow{6}{*}{31.4}  \\
 & $\tilde{\ell}_L^\pm\tilde{\nu}_\ell$   &  $5\ell\,0\tau,\,\,3\ell\,2\tau,\,\,1\ell\,4\tau$ & & \\
 & $\tilde{\ell}_L^\pm\tilde{\ell}_L^\mp$ &  $6\ell\,0\tau,\,\,4\ell\,2\tau,\,\,2\ell\,4\tau$&  & \\
  & $\tilde{\nu}_\tau\tilde{\nu}_\tau$  & $0\ell\,2\tau\,(+2 W)$ & &\\
 & $\tilde{\tau}_2^\pm \tilde{\nu}_\tau$  &  $0\ell\,2\tau\,(+W+Z)$ & &\\
 & $\tilde{\tau}_2^\pm\tilde{\tau}_2^\mp$   &  $0\ell\,2\tau\,(+2Z)$& & \\
 \hline
\multirow{6}{*}{Slepton-bino} & $\tilde{\nu}_\ell\tilde{\nu}_\ell$  & $4\ell\,0\tau,\,\,2\ell\,2\tau,\,\,0\ell\,4\tau$ & \multirow{6}{*}{51.5}  & \multirow{6}{*}{43.0}  \\
 & $\tilde{\ell}_L^\pm\tilde{\nu}_\ell$   &  $5\ell\,0\tau,\,\,3\ell\,2\tau,\,\,1\ell\,4\tau$& &\\
  & $\tilde{\ell}_L^\pm\tilde{\ell}_L^\mp$ &  $6\ell\,0\tau,\,\,4\ell\,2\tau,\,\,2\ell\,4\tau$ & & \\
 & $\tilde{\nu}_\tau\tilde{\nu}_\tau$  & $4\ell\,0\tau,\,\,2\ell\,2\tau,\,\,0\ell\,4\tau$& &\\
 & $\tilde{\tau}_2^\pm \tilde{\nu}_\tau$  &  $4\ell\,1\tau,\,\,2\ell\,3\tau,\,\,0\ell\,5\tau$ & &\\
 & $\tilde{\tau}_2^\pm\tilde{\tau}_2^\mp$   &  $4\ell\,2\tau,\,\,2\ell\,4\tau,\,\,0\ell\,6\tau$& & \\
 \hline
\end{tabular}
\caption{Production modes and final state topologies, for the flavor-democratic benchmark scenarios.  In the final two columns, we show the fraction of events for each benchmark that contain same-sign dileptons or trileptons, inclusively, for $m_{NLSP}\approx 150$ GeV and $m_{prod}\approx$ 200~GeV\@. Here the leptons (meaning, $e$ or $\mu$) are required to satisfy $|\eta|<1$ and $p_T>10$~GeV, and are simulated with the default PGS4 CDF simulation. Note that the benchmarks with slepton production have topologies with more leptons than the benchmarks with -ino production, resulting in a higher branching fraction to same-sign dileptons and trileptons.}
\label{tab:topologies}
\end{center}
\end{table}

We comment that our strategy of choosing simple spectral types is distinct from the approach, advocated elsewhere~\cite{Dube:2008kf}, of enumerating the limits on all possible final state topologies (determined by the final state multiplicities and kinematics). ÊThis is because even the simplest possible spectra can lead to many final state topologies at once, with different branching fractions. ÊOur four benchmark scenarios together contain 32 distinct final state topologies, and it would be cumbersome to present separate limits for each.

\begin{figure}[!t]
\begin{center}
\includegraphics[scale=0.57]{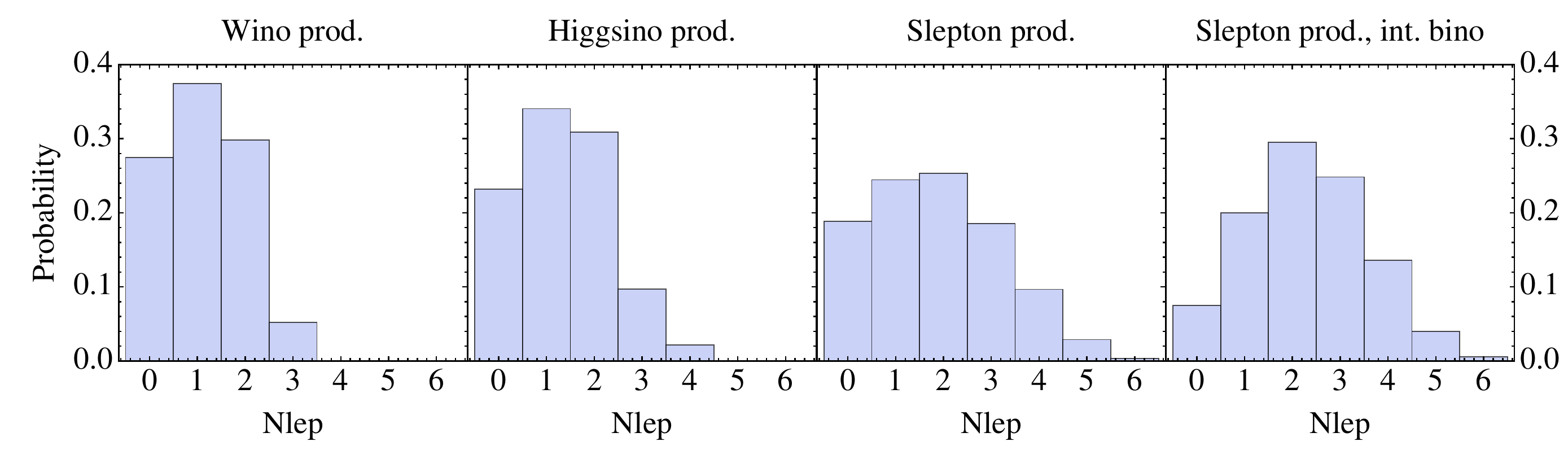} 
\caption{Number of leptons per event for each benchmark, simulated with PGS and requiring $|\eta|<1$, $p_T>10$ GeV\@.}\label{fig:nlep}
\end{center}
\end{figure}

Figure~\ref{fig:nlep} contains histograms of the number of leptons produced for each benchmark, subject to the same cuts, and for the same choice of parameters.  We see that for wino and higgsino production, the fraction of events with more than two leptons is much smaller than for slepton production. We also notice that for slepton production, events with 4 or 5 leptons are not uncommon. (Including heavier states in the SUSY cascade, e.g.\ winos or higgsinos, would lead to even more leptons, up to 8 in all.) Finally, we learn that for slepton production without intermediate bino there are many more zero lepton events compared to slepton production with intermediate bino. This is a sign of the lepton degradation in the third generation production modes, as discussed above.

\begin{figure}[!t]
\begin{center}
\includegraphics[scale=.57]{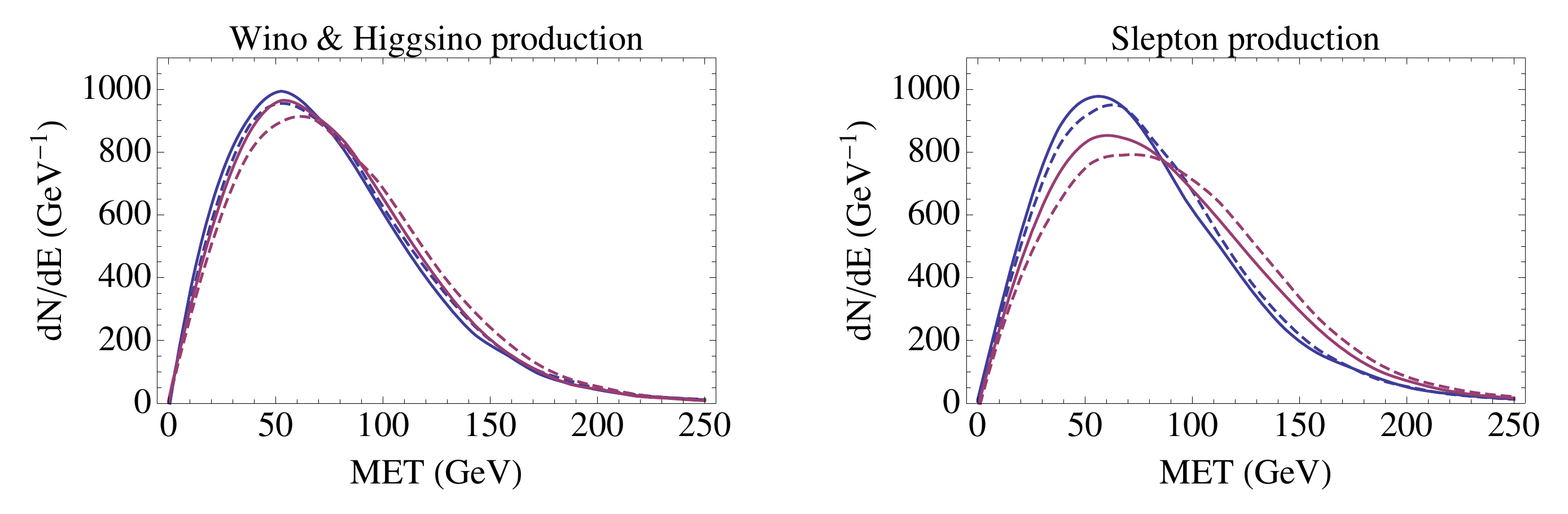} 
\caption{$\met$ distributions for the four benchmark scenarios, with 100k events. In all cases, we have taken $m_{prod}=200$~GeV\@. The blue (red) curves are for $m_{\tilde\ell_R}=100$ GeV ($m_{\tilde\ell_R}=150$~GeV). The solid curves in the left and right plots correspond to wino and slepton production, respectively, and the dashed curves correspond to higgsino production and slepton production with intermediate bino, respectively. We see that all four benchmarks have significant $\met\gtrsim 50$~GeV\@.}
\label{fig:met}
\end{center}
\end{figure}

In fig.\ \ref{fig:met}, we show the missing energy distributions (again simulated using PGS) for the four benchmark scenarios, and two choices of the $m_{\tilde\ell_R}$ mass.  We see that models with slepton co-NLSP, with masses in the range of interest, are characterized by significant $\met \gtrsim 50$~GeV\@.\footnote{ We also learn that for higgsino and wino production, the $\met$ distribution is largely independent of $m_{\tilde\ell_R}$. This surprising fact is because in these cases, the $\met$ comes mainly from the (vector) sum of the gravitino momenta {\it plus} the neutrino momentum coming from $\chi_1^\pm \to \tilde\tau_1^\pm + \nu_\tau$. Since the neutrino momentum is $\sim m_{prod}-m_{\tilde\ell_R}$ while the gravitino momenta are $\sim m_{\tilde\ell_R}/2$ each, the total $\met$ is $\sim m_{prod}$ on average, i.e.\ the right-handed slepton mass cancels out.  For the slepton production modes, there are also neutrinos from the decays of the left-handed sleptons, but they do not take up the entire energy, so the cancellation is less prominent.} This should be compared with the typical $\met$ distribution in the mSUGRA scenario (see appendix \ref{sec:GGMvSugra}), which is generally much softer.

\begin{figure}[!t]
\begin{center}
\includegraphics[scale=0.75]{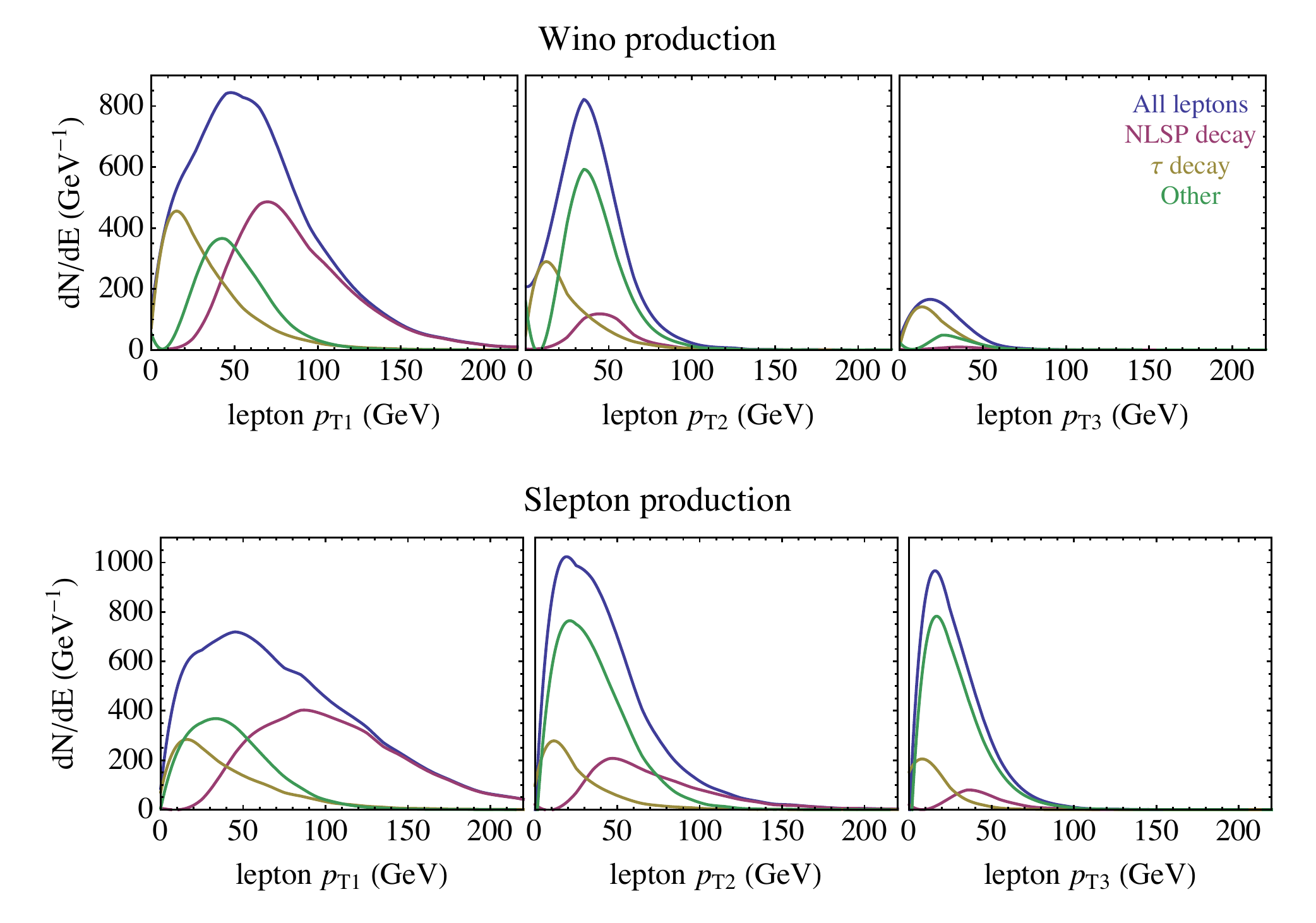} 
\caption{The three hardest lepton $p_T$'s for the wino and slepton benchmarks, with 100k events simulated with PGS and requiring $|\eta|<1$.  We have fixed $m_{\tilde \ell_R}=150$~GeV and $m_{\tilde W}$ (or $m_{\tilde\ell_L}$) $=200$ GeV\@.  The red curve represents leptons coming from $\tilde e_R$ and $\tilde \mu_R$ decays. The yellow curve is leptons coming from $\tau$ decays. The green curve is leptons coming from decays of heavier states in the SUSY cascade. The blue curve represents the sum of all these curves.}
\label{fig:allleppt}
\end{center}
\end{figure}

Finally, in fig.\ \ref{fig:allleppt}, we show the $p_T$'s of the three hardest leptons, for the wino and slepton benchmarks, again for the same choice of parameters.  The higgsino and slepton with bino benchmarks are not shown, but the distributions look similar to the wino and slepton benchmarks, respectively.  We have again taken $m_{\tilde\ell_R}=150$ GeV and $m_{prod}=200$ GeV\@. We have broken down the distributions into leptons coming from $\tilde e_R$, $\tilde\mu_R$ decay (red), leptons coming from $\tau$ decays (yellow), and leptons coming from decays of heavier states (green). We see that the hardest lepton tends to be quite energetic $p_T\gtrsim 50$ GeV, and mostly composed of direct right-handed slepton decays. The composition shifts towards $\tau$ decays and decays of heavier states as one moves to the second and third hardest leptons. We note that the third hardest lepton is much more energetic and numerous in the slepton production scenario, compared with the wino scenario.

\section{Overview of Tevatron Searches}\label{sec:searches}

We now review the existing Tevatron searches which are most relevant for GMSB with slepton co-NLSPs: same-sign dileptons$+\met$ and trileptons$+\met$. We will use these searches to determine the current limits and future reach on slepton co-NLSPs in section~\ref{sec:results}.

\subsection{SS Dileptons}

The simplest leptonic channel with low SM background is two leptons of the same charge.  Same-sign dileptons have been searched for at CDF~\cite{Abulencia:2007rd}  and D0~\cite{D0Dilep}, both with 1~fb$^{-1}$.  The CDF search uses an inclusive approach and includes the $ee$, $e \mu$ and $\mu \mu$ channels.  The D0 search includes only the $\mu \mu$ channel, and the cuts are not inclusive but are instead tuned to a particular mSUGRA-type model point. 
For this reason we choose to focus here on the CDF search.  We also comment that CDF has conducted another same-sign dilepton search with 2.7~fb$^{-1}$~\cite{Aaltonen:2009nr}.  However, motivated by fourth generation quarks, they look in the $l^\pm l^\pm b j \met$ channel, in particular requiring at least two jets, one of which must be tagged as a $b$-jet.  This search is therefore not useful for constraining our benchmark scenarios and we will not consider it further.\footnote{Figure~1 of Ref.~\cite{Aaltonen:2009nr} does show the missing energy distribution of same-sign dilepton events without requiring jets in the event.  However, the uncertainties here are $\mathcal{O}(100\%)$, so this figure does not appear to be useful for setting limits.}

The CDF search employs a simple set of inclusive cuts on like-sign dilepton pairs, shown in table~\ref{tab:SScuts}. With these cuts, 44 events are observed in the data, with an expected background of $33.2\pm 4.7$.
The largest backgrounds are Drell-Yan plus an untagged photon conversion and leptonic decays of dibosons, $WZ$ and $ZZ$.  Taus are not specifically tagged, but the search is sensitive to leptonically decaying taus, which account for 35\% of tau decays.  The strength of this search is its inclusive approach, which allows for a straightforward interpretation of its limits on GMSB\@.

\begin{table}[!t]
\begin{center} \begin{tabular}{| c |}
\hline
SS Dilepton Cuts\\
\hline
$p_T^1, p_T^2> 20, 10$~GeV \\
$|\eta_{1,2}|<1$ \\ 
$m_{12}>25$~GeV \\
\hline
\end{tabular}\end{center}
\caption{The cuts employed by CDF for the 1~fb$^{-1}$ same-sign dilepton search~\cite{Abulencia:2007rd}.
\label{tab:SScuts}}
\end{table}

The CDF paper provides detailed kinematic distributions, such as $\met$ and lepton $p_T$'s. These allow us to optimize somewhat for slepton co-NLSPs. The simplest option is to augment the inclusive CDF cuts with a hard MET cut of $\met > 60$~GeV\@.  This cut alone reduces the data to 4 events and the background to 2.2 events, while leaving most of the GMSB signal (as can be seen from fig.\ \ref{fig:met}).  In the rest of the paper, we will use this selection to set limits and infer reach for slepton co-NLSPs.

\subsection{Trileptons}

Both Tevatron experiments have also searched for new physics in the trilepton channel, which has low SM background.  The most recent CDF results \cite{CDFTrilep3p2} use 3.2 fb$^{-1}$ of data; these are a straightforward update of an earlier 2~fb$^{-1}$ analysis~\cite{Aaltonen:2008pv}. Meanwhile, the latest D0 results~\cite{Abazov:2009zi} use 2.3~fb$^{-1}$.  The CDF search uses a cut-based analysis that is simple to reproduce with PGS\@.  On the other hand, the D0 search uses a neural network to tag hadronic $\tau$ decays, and this makes it more difficult to simulate the search in order to interpret the limits on GMSB\@.  For this reason, and because of the larger luminosity, we choose to focus on the CDF search of Ref.~\cite{Abazov:2009zi}.

We now give a simplified description of the CDF trilepton search, focusing on the cuts that are important for understanding the limit on slepton co-NLSPs.  For a full description of the analysis, we refer the reader to Refs.~\cite{Aaltonen:2008pv, Abazov:2009zi} and especially the thesis \cite{SourabhThesis}.  The CDF trilepton search selects events from two categories: three leptons, $lll$, or two leptons and one isolated track $llT$\footnote{Leptons are further broken down into {\it tight} and {\it loose} categories, where the {\it tight} leptons pass stricter ID requirements.  The distinction between these lepton types will not be very important here, and for our simulation we sum the tight and loose bins, as discussed in appendix~\ref{app:calibratePGS}.}.  The latter category is designed to catch 1-prong hadronic $\tau$ decays, and this significantly improves the acceptance for models with $\tau$'s in the final state, albeit with higher backgrounds.  The cuts of the search are listed in table~\ref{tab:Trilepcuts}.  The trileptons must satisfy the inclusive cuts shown to the left, and events are vetoed if they satisfy any of the criteria on the right.

\begin{table}[!t]
\begin{center} \begin{tabular}{| c | c || c | c |}
\hline
 \multicolumn{2}{|c||}{Inclusive Cuts} & \multicolumn{2}{c|}{Non-Inclusive Cuts (Vetoes)}  \\
\hline
$lll$ & $p_T^1, p_T^2, p_T^3 > 15, 5 ,5$~GeV & Charge & $\Sigma Q = \pm 3$ \\
\hline
$llT$ & $p_T^1, p_T^2, p_T^t > 15, 5 ,10$~GeV & $Z$ & $76 < m_{l^+ l^-} < 106$~GeV \\
\hline
\multicolumn{2}{|c||}{$|\eta| \lesssim 1$} & 4th lepton & $p_T>10$~GeV and $|\eta|\lesssim1$ \\
\cline{3-4}
\multicolumn{2}{|c||}{$\met > 20$~GeV}  & 2nd jet & $E_T> 15$~GeV and $|\eta|<2.5$ \\
\hline
\end{tabular}\end{center}
\caption{The cuts employed by CDF for the 3.2~fb$^{-1}$ trilepton search~\cite{CDFTrilep3p2}.  The inclusive cuts on the left select events with three leptons or two leptons and a track.  Then, events are vetoed if the lepton charge sums to $\pm3$, if the leptons reconstruct the $Z$, or if there is a fourth lepton or at least two jets.
\label{tab:Trilepcuts}}
\end{table}

After cuts, one event is observed in the $lll$ category, with $1.5\pm 0.2$ expected; and 6 events are observed in the $llT$ category with $9.4\pm 1.4$ expected. The main backgrounds to this search include leptonic decays of dibosons $WZ / ZZ$ and $t \bar t$, and Drell-Yan accompanied by a photon conversion, isolated track, or fake lepton.  The low background rates are why the trilepton channel places sensitive limits on new physics with lepton-rich final states. 

Although we have chosen to focus here on the CDF trilepton search, we comment that the D0 search may be more sensitive to some GMSB models.  It has no jet veto for some of its search bins, and it allows up to two hadronic 1-prong taus, which means it probably  has better reach for GMSB models with colored production or with $\tau$-rich decays.  We will revisit this issue in section \ref{sec:future}.

\section{Results}\label{sec:results}

In this section, we present the current Tevatron limits, 95\% exclusion reach, and discovery potential for the four benchmark slepton co-NLSP scenarios described in section~\ref{sec:benchmarks}.  As discussed in section \ref{sec:searches}, we will focus on the CDF trilepton and same-sign dilepton searches, as these should provide the best sensitivity to slepton co-NLSPs.  (Keep in mind, for the latter we are including a $\met>60$ GeV cut as inferred from the plot of $\met$ in~\cite{Abulencia:2007rd}.) Since these searches did not discuss slepton co-NLSPs, we have estimated on our own, using standard Monte Carlo tools, the signal acceptances for the various analyses. For more details on our simulations, see appendix \ref{app:calibratePGS}.

\subsection{Current Limits and Estimated Exclusion Reach}

\begin{figure}[!t]
\begin{center}
\includegraphics[scale=0.5]{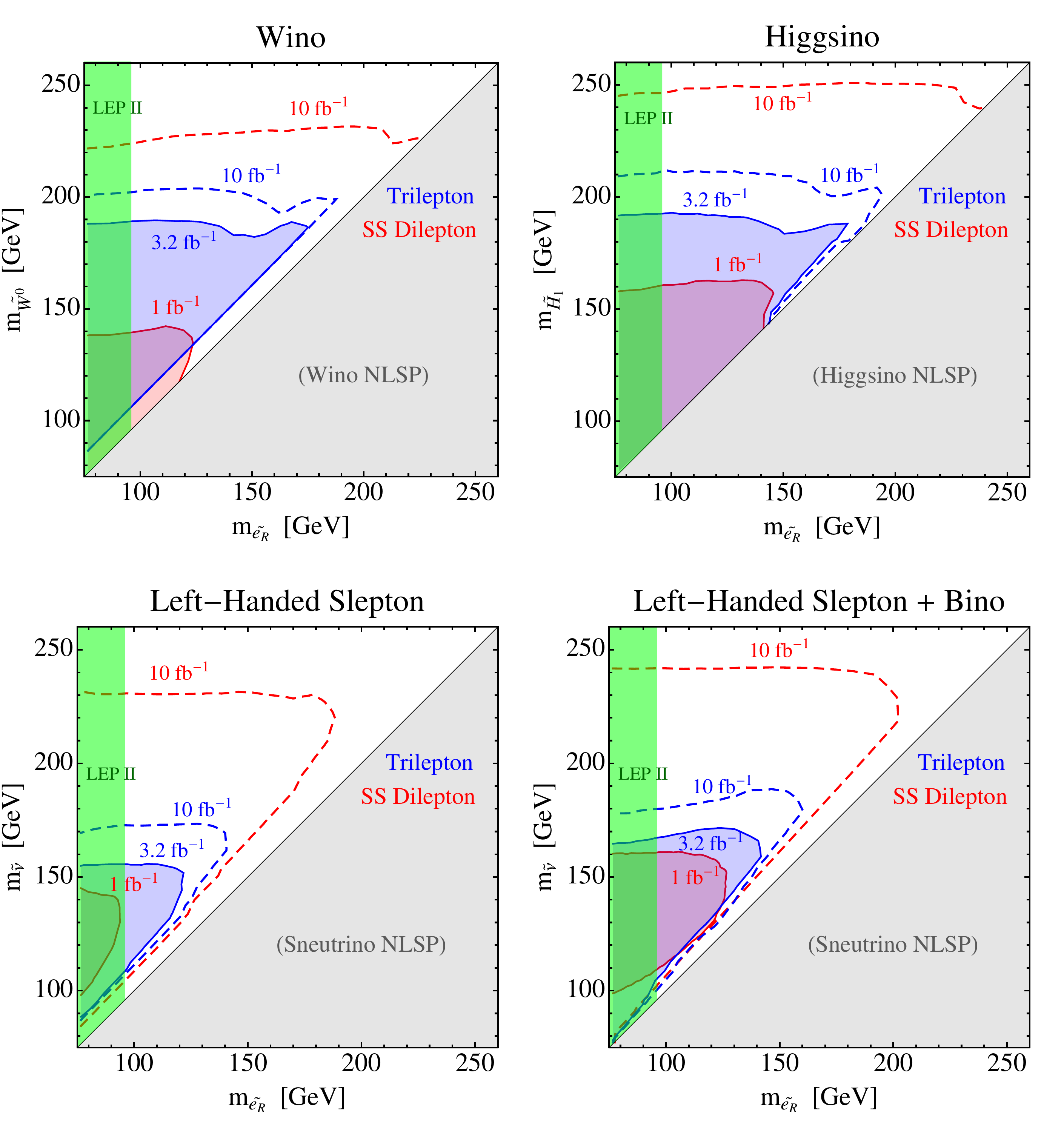}
\caption{The current (filled) and projected (dashed) 95\% exclusion limits, from the trilepton~\cite{CDFTrilep3p2} (blue) and same-sign dilepton~\cite{Abulencia:2007rd} (red) searches, on the four benchmark scenarios of table~\ref{tab:bench}.
To calculate the limits, we have used the CL$_s$ statistic~\cite{Junk:1999kv}, which is designed to limit the impact from downward fluctuations of the background.  We also show the limit on the right-handed slepton masses (green band), as set by LEP2~\cite{LEP2}.
}
\label{fig:limits}
\end{center}
\end{figure}

The current 95\% limits on our four benchmark slepton co-NLSP scenarios are shown in fig.~\ref{fig:limits}. In addition, we show the expected limit that could be set with 10 fb$^{-1}$ of data. Since this plot is among the main results of our paper, let us take a moment and highlight some of its features.

For each of the four benchmarks, we find that the limit depends primarily on the mass of the heavier state that is being produced (wino, higgsino, or left-handed slepton). This is consistent with the fact that
the production cross-section is a steep function of mass, so ${\mathcal O}(1)$ variations in the signal acceptance do not result in big changes in the exclusion contour. We see that for the existing limits, the constraint from trileptons is always stronger because it enjoys a larger luminosity (3.2 fb$^{-1}$) than the latest dilepton search (1 fb$^{-1}$). However, the projected limits at 10 fb$^{-1}$ show dileptons winning out decisively. 

Of course, this is not an entirely fair comparison, since we have used the $\met$ distribution in \cite{Abulencia:2007rd} to slightly optimize SS dileptons for slepton co-NLSPs.  We believe that a similar improvement is possible for trileptons if the cuts are optimized (for our suggestions in this direction, see section~\ref{sec:discussion}).  
To emphasize this point, we note that the difference in reach between the trilepton and same-sign dilepton searches is largest for the two left-handed slepton production benchmarks.  This is because left-handed sleptons lead to higher lepton multiplicities, where the same-sign dilepton search benefits from its inclusive approach, and where the trilepton search loses out by vetoing on additional central leptons (among other things).

It might seem surprising that the projected 10 fb$^{-1}$ limit on left-handed sleptons is so strong, since their cross-section is a factor of 5-10 smaller than the wino or higgsino production rates, as shown in figure~\ref{fig:prod}.  But again, this is because left-handed slepton production leads to more leptons in the final state (see table~\ref{tab:topologies} and figure~\ref{fig:nlep}), and therefore a higher experimental acceptance from SS dileptons. The signal acceptances of the various searches, calculated at the boundaries of the excluded regions, are summarized in table \ref{tab:acceptance}.

\begin{table}[t!]
\begin{center} \begin{tabular}{| c | c | c | c  | c | c |}
\hline
\multirow{2}{*}{channel}&\multirow{2}{*}{$m_{\rm prod}$~[GeV]}  & \multirow{2}{*}{$\sigma$~[fb]}& \multicolumn{2}{c|}{trileptons} & SS dilepton\\
\cline{4-6}
 & &  &  $lll$ & $llT$& $l^\pm l^\pm$ \\
\hline
wino & 190 & 69 & $8\times10^{-3}$ & .02  & .04 \\
higgsino & 190 & 47 & .02 & .02  & .08 \\
slepton & 155 & 41 &.02 & .01 & .11 \\
slepton +bino & 170 & 23 & .04 & .03 & .24 \\
\hline
\end{tabular}\end{center}
\caption{The masses, cross-sections, and acceptances at the edge of the current 95\% limits for the four benchmark scenarios.  The cross-sections are leading order, and in each case we fix $m_{\tilde e_R}= 120$~GeV\@. 
 \label{tab:acceptance}}
\end{table}

The experimental sensitivity of the trilepton and SS dilepton searches can degrade when the mass splitting between the production mode and right-handed sleptons is small. The reasons for this vary between the different searches and the different benchmark scenarios. For winos, the trilepton search has a squeezed region, simply because the third lepton $p_T$ is being squeezed out. The dilepton search has no analogous squeezed region, since there the two SS leptons can come from the decay of the co-NLSPs.  For higgsino production, the trilepton and dilepton searches have no squeezed region. This is because there is always a sizeable mass splitting between the two neutral higgsinos, $m_{\tilde H_2} - m_{\tilde H_1} \sim 10$~GeV, so the third lepton can come from the decay of $m_{\tilde H_2}$ to the co-NLSPs. Finally, for the two slepton production scenarios, both the trilepton or dilepton searches have a squeezed region. Here the reason is because the sneutrinos decay to right-handed sleptons through a three-body decay in the squeezed regime, but for low-scale gauge mediation, this three-body decay can lose to direct sneutrino decay to gravitino, $\tilde \nu \rightarrow \nu+ \tilde G$.  Therefore, in the squeezed regime, the sneutrinos become co-NLSPs and there is not enough branching ratio into trilepton or like-sign dilepton final states to produce a limit.

\subsection{Estimated Discovery Potential}

\begin{figure}[t!]
\begin{center}
\includegraphics[scale=0.45]{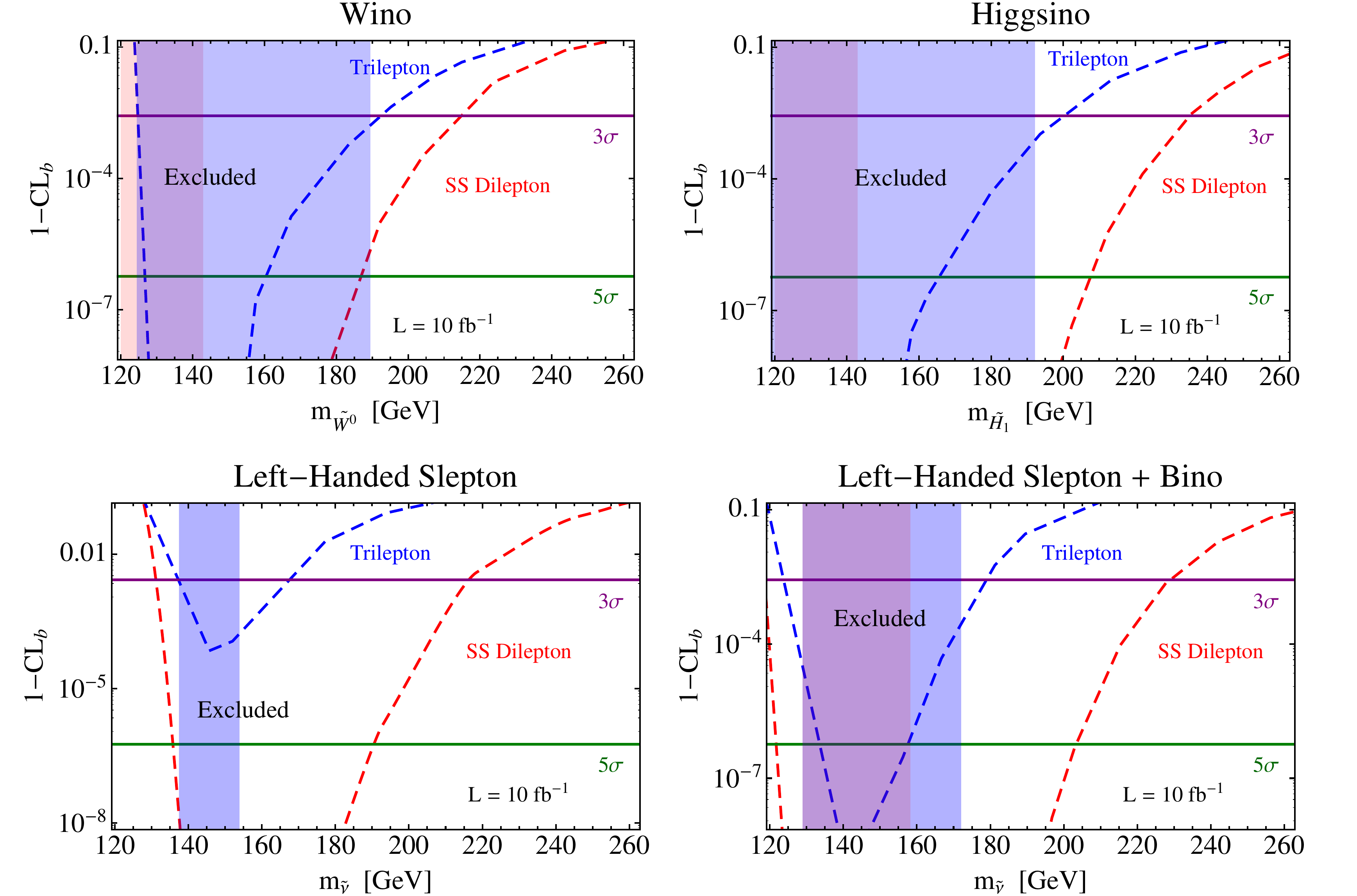}
\caption{The 10~fb$^{-1}$ discovery potential of the  trilepton~\cite{CDFTrilep3p2} (blue) and same-sign dilepton~\cite{Abulencia:2007rd} (red) searches, fixing $m_{\tilde e_R} = 120$~GeV\@.  We use the CL$_b$ statistic~\cite{Junk:1999kv} which measures the probability of the background fluctuating to give the signal plus background.  The green (purple) lines show the thresholds for $5\sigma$ ($3\sigma$) discovery and the blue (red) shaded region is already excluded at 95\% by the existing trilepton (dilepton) search.  For all four benchmarks we find that both searches can discover supersymmetry at 3$\sigma$ for a range of allowed parameters, and the SS dilepton search has the potential for $5\sigma$ discovery for all but wino production. 
\label{fig:discover}}
\end{center}
\end{figure}

Finally, let us also estimate the potential for $3 \sigma$ and $5\sigma$ discovery in 10~fb$^{-1}$. Here we specialize to a model line, so that we can show the confidence level statistic in more detail. The result is shown in figure~\ref{fig:discover}, fixing $m_{\tilde e_R} = 120$~GeV\@.  For each channel, we find a significant range of parameters where $3\sigma$ evidence is possible with the same-sign dilepton search.  Specifically, $3\sigma$ evidence is possible up to 215~GeV for winos, 235 GeV for higgsinos, and 230 (220) GeV for sleptons with (without) an intermediate bino.  There is even a significant range of left-handed slepton masses consistent with the current limit (about 30 GeV in mass), where a $5\sigma$ discovery is possible at the Tevatron!  
It is exciting to see that such a dramatic discovery is still possible at the Tevatron.  This significant discovery potential is possible because the current searches have not been optimized for the slepton co-NLSP type signature, and because the same-sign dilepton search has not been updated since the relatively small luminosity of 1~fb$^{-1}$.

\section{Beyond Our Benchmarks}\label{sec:future}

\subsection {Tau Rich Decays}

Above, we have explored the Tevatron limits and reach for gauge mediation models with flavor democratic decays. This is because the existing Tevatron searches for multilepton events have the best sensitivity in this case. But as discussed in section~\ref{sec:taxonomy}, it is also a very common situation to have decays producing mostly $\tau$s.  This can happen either with slepton co-NLSPs but preferential decays to $\tilde\tau_1$; or with stau NLSPs, where all SUSY cascades necessarily end in $\tilde\tau_1$.

Such models are obviously much less constrained at the Tevatron than models with slepton co-NLSP and flavor democratic decays.  The existing Tevatron searches are mainly sensitive to these events when the $\tau$s decay leptonically (which occurs with a 35\% branching fraction). Since the leptons arise from a three-body decay $\tau\to \ell+\nu_\ell+\nu_\tau$, the daughter lepton is significantly softer than the mother $\tau$.  The poor experimental acceptance for final states with $\tau$'s arises from a combination of branching ratios and softer leptons. Note that the trilepton search will also be sensitive to events with one-prong hadronic $\tau$ (50\% branching fraction), provided the other $\tau$'s decay leptonically. Here however the backgrounds are much higher.

\begin{figure}[t!]
\begin{center} \includegraphics[scale=0.55]{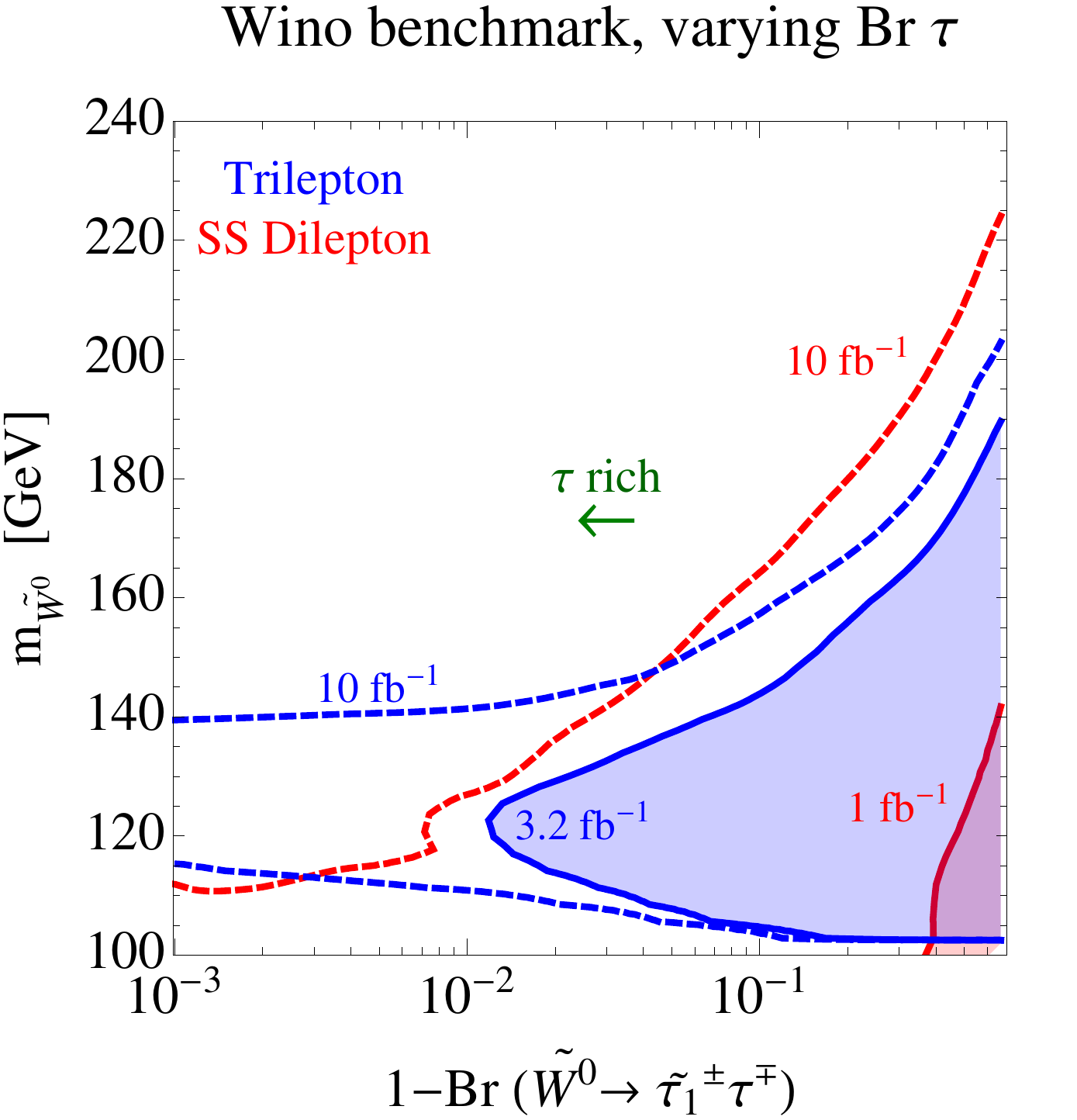} \end{center}
\caption{The current limit (solid) and future reach (dashed)  from the CDF same-sign dilepton and trilepton searches, for wino production with varying branching fraction to $\tilde\tau_1$ and wino mass. For this plot, the right-handed slepton masses were fixed at the LEP II limit, namely 96 GeV\@. The right-hand side of the plot corresponds to the flavor democratic case studied above. Moving leftward on the plot increases the fraction of events populated with $\tau$s, worsening the experimental sensitivity.
\label{fig:TauRich}}
\end{figure}

To illustrate how the sensitivity of the Tevatron searches degrades when models are tau rich, we show the 95\% limit and reach contours on figure~\ref{fig:TauRich}, for a simple deformation of the  wino benchmark scenario defined in section~\ref{sec:taxonomy}. Recall that in the original benchmark, $\tan \beta = 2$ and $M_1=300$ GeV, and so the neutral wino decays democratically to $\tilde \tau_1$, $\tilde e_R$, and $\tilde \mu_R$. In fig.\ \ref{fig:TauRich}, we have considered a modified scenario where  ${\rm Br}(\tilde W^0 \to\tilde\tau_1+\tau)$ is allowed to vary. (One can think of this as varying $\tan\beta$ and $M_1$.) As can be seen from the figure, the reach and limit are significantly degraded as ${\rm Br}(\tilde W^0 \to\tilde\tau_1+\tau)\to 1$. {\it In fact, there is currently no limit whatsoever from Tevatron searches for the fully $\tau$-rich case.} The trilepton search loses sensitivity at lower values of $m_{\tilde W_0}$, because the third lepton, which must come from wino decays, becomes too soft. However, at higher values of the wino mass, trileptons have the best reach, because they include the $llT$ category. This should be contrasted with the flavor democratic case, where SS dileptons have the best reach, because of its more inclusive  nature.

The lack of any current Tevatron limit on tau-rich scenarios is a huge gap in the search for GMSB\@. This may present an opportunity for early LHC running if dedicated searches are developed for tau-rich final states.  More generally, the results presented in this subsection highlight the importance of developing search strategies that are more sensitive to hadronic taus. We conclude this section with an important caveat to the above discussion.  We have not studied the limits resulting from the D0 trilepton search~\cite{Abazov:2009zi}, which was conducted with 2.3~fb$^{-1}$ and includes a search category with a muon and two (hadronic 1-prong) taus.  This bin may improve on the limit presented in figure~\ref{fig:TauRich}.  One difficulty for deriving such a limit is that the D0 search relies on a neural network for tagging hadronic tau decays, which is difficult to replicate using PGS\@.

\subsection{A Look Towards the LHC}
\label{sec:LHC}

In this paper, we have focused on the limits and reach, resulting from electroweak production at the Tevatron.  
The next step will be to search for gauge mediation with slepton co-NLSPs at the LHC (see for example the multilepton signatures discussed in~\cite{Baer:2000pe, DeSimone:2008gm,DeSimone:2009ws}), which is now running with center of mass energy of 7~TeV\@.  In order to understand how well the LHC can compete with the Tevatron, we can compare the ratios of cross-sections of interest.  For example, 200 GeV left-handed sleptons  satisfy $\sigma_{LHC} / \sigma_{TeV} \simeq {\mathcal O}(10)$.  More generally, the Tevatron with $\sim$ 10~fb$^{-1}$ will produce at least as many light uncolored states as the LHC with $\lesssim 1~\mathrm{fb}^{-1}$.  Meanwhile, many of the backgrounds will be larger at the LHC, implying that the Tevatron retains an advantage for electroweak production~\cite{Campbell:2010ff}.  On the other hand, the early LHC enjoys a significant advantage for $gg$ initiated processes, and 450 GeV gluinos satisfy $\sigma_{LHC} / \sigma_{TeV} \simeq 500$, where we have fixed the squark masses to 1~TeV\@.  Colored SUSY production at the LHC will very quickly overtake the Tevatron, providing natural channels to look for GMSB in the first run of the LHC\@.  We stress that the early LHC and the Tevatron are largely {\it complementary} experiments from the point of view of SUSY discovery -- the former is well-suited for light colored states, while the latter (owing to its large integrated luminosity and lower QCD backgrounds) is ideal for light EW states.

We will address colored production of slepton co-NLSPs at the early LHC in an upcoming publication~\cite{JoshDavidLHC}.  As a preliminary step, we here establish the limit from the Tevatron for an example process.
We consider slepton co-NLSPs with a low energy spectrum consisting of a light gluino, $\tilde g$, flavor-degenerate right-handed sleptons, $\tilde l_R$, and an intermediate bino, $m_{\tilde g} > m_{\tilde B} > m_{\tilde l_R}$.\footnote{Keep in mind that such a spectrum is perfectly allowed in GGM, but never occurs in MGM where the gluino is always much heavier than the bino and the right-handed sleptons.} The gluino decays to the bino through a three-body decay mediated by an off-shell squark,
\begin{equation}
\tilde g \to \tilde B + 2 j
\end{equation}
and then the bino decays democratically to the right handed sleptons, as in our fourth benchmark above, 
\begin{equation}
\label{eq:BinoDecay}
\tilde B \to (\tilde e_R^\pm,e^\mp),\,\,\,(\tilde\mu_R^\pm,\mu^\mp),\,\,\,(\tilde\tau_1^\pm,\tau^\mp)
\end{equation}
with equal branching ratios ($1/6$ per final state).  And as usual, the slepton co-NLSPs decay to their superpartner and a gravitino.  The resulting final state is $4 l + 2 j + E\!\!\!\!/_T$.\footnote{A similar final state is produced by a spectrum with the bino lifted, $m_{\tilde B} > m_{\tilde g}$, in which case the gluino's dominant decay is 4-body, but we will not consider this situation here.}
For this scenario, we choose a low SUSY breaking scale with a gravitino mass of $m_{\tilde G} = 1$~eV, and we fix the bino mass to be halfway between the gluino and right-handed sleptons, $m_{\tilde B} = (m_{\tilde g}+m_{\tilde l_R})/2$. 
When determining the gluino production cross-section, we assume degenerate squarks of mass $m_{\tilde q} = 1$~TeV, and we use the NLO value of Prospino~\cite{Beenakker:1996ed,Beenakker:1996ch}.

\begin{table}[t!]
\begin{center} \begin{tabular}{ | c | c | c  | c | c |}
\hline
\multirow{2}{*}{$m_{\rm \tilde g}$~[GeV] } & \multirow{2}{*}{$\sigma$~[fb] }& \multicolumn{2}{c|}{trileptons} & SS dilepton\\
\cline{3-5}
& &  $lll$ & $llT$& $l^\pm l^\pm$ \\
\hline
400 & 36 & $10^{-4}$ & $4\times10^{-5}$ & .19 \\
\hline
\end{tabular}\end{center}
\caption{The gluino mass, production cross-section, and acceptance at the edge of the current 95\% limit.  We fix $m_{\tilde e_R}=120$~GeV, and the cross-section is the NLO value from Prospino~\cite{Beenakker:1996ed,Beenakker:1996ch}. \label{tab:GluinoAcceptance}}
\end{table}

\begin{figure}[t!]
\begin{center} \includegraphics[scale=0.55]{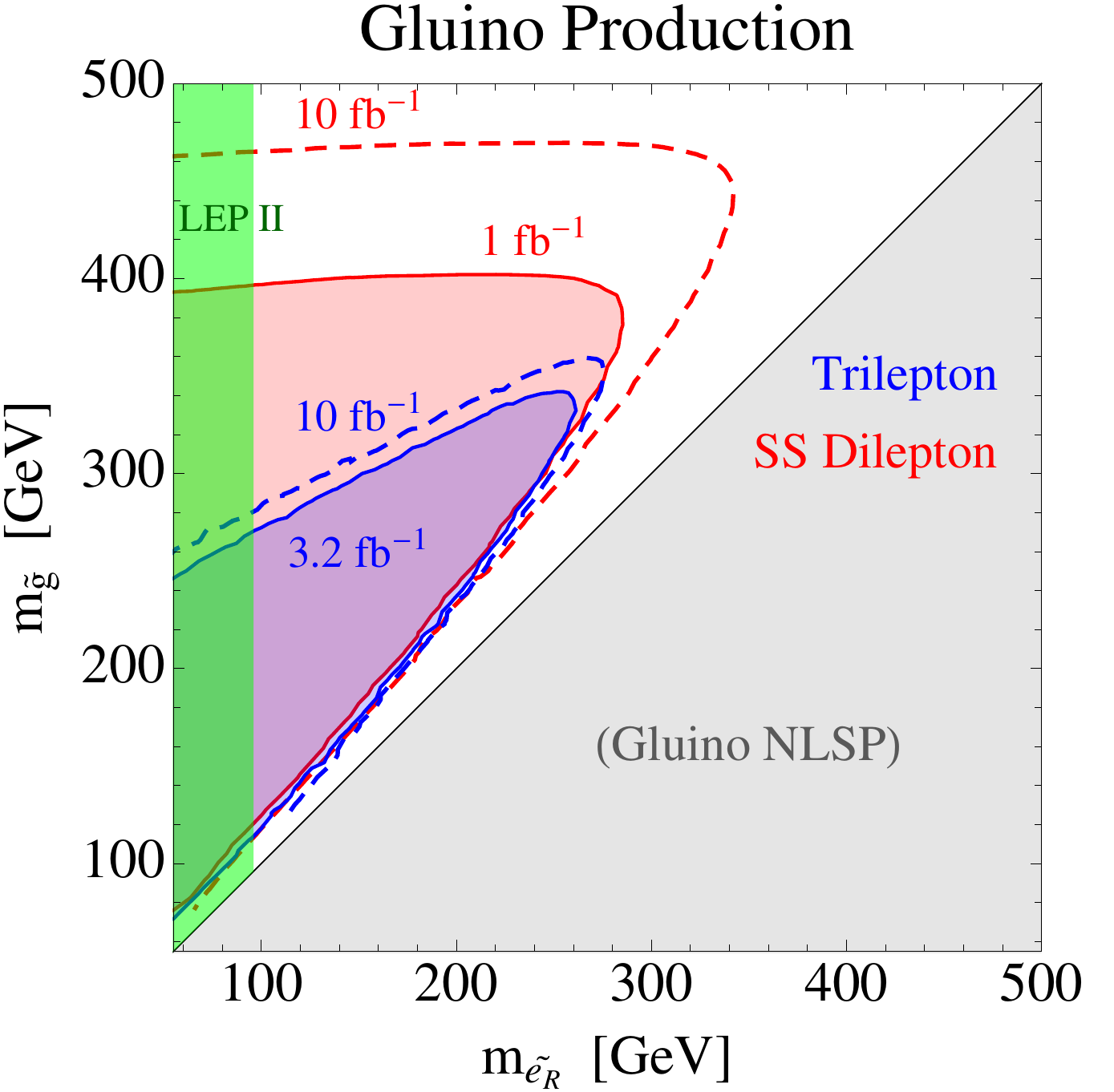} \end{center}
\caption{The current limit (solid) and future reach (dashed) for gluino production with slepton co-NLSPs.  We use the CDF same-sign dilepton and trilepton searches, and choose a benchmark scenario with an intermediate bino with mass halfway between the gluino and right-handed sleptons.  The trilepton search includes a jet veto and therefore produces a weaker limit than the same-sign dilepton search, which is inclusive.
\label{fig:gluino}}
\end{figure}

We present the current limits from the CDF same-sign dilepton and trilepton searches, along with the 10~fb$^{-1}$ reach, in figure~\ref{fig:gluino}.   In table~\ref{tab:GluinoAcceptance}, we show the gluino mass, production cross-section, and experimental acceptances at the edge of the current 95\% exclusion region.  We see that the dilepton search has an acceptance that is $10^3-10^4$ times larger than the trilepton search.  This is because the CDF trilepton search includes a jet veto that throws out most of the signal (this was designed to reduce the background from $t\bar t$). Consequently, we find that the same-sign dilepton search sets a significantly stronger limit than the CDF trilepton search, despite the smaller luminosity of  1~fb$^{-1}$ versus 3.2~fb$^{-1}$. 
We comment that the D0 trilepton search~\cite{Abazov:2009zi} did not use a jet veto for some of their search bins, and we therefore expect this search to set a stronger limit that the CDF trilepton search, for this scenario.  We do not derive this limit here.  It would also be interesting to infer a limit from the CDF search for $l^\pm l^\pm bj \met$~\cite{Aaltonen:2009nr}, although their requirement for a jet with a $b$-tag makes them sensitive only to SUSY cascades involving heavy flavor squarks.

\section{Discussion}\label{sec:discussion}

\subsection{How to Optimize for GMSB}

In this section, we discuss how the above searches can be optimized for the slepton co-NLSP signal, and we suggest a few additional searches that could improve the Tevatron's sensitivity for GMSB\@.  Our suggestions here are mostly culled from the preceding analysis. Some are rather obvious, but we hope it will be useful to collect them all here in one place.

Many of the SUSY searches at the Tevatron, such as the trilepton search described above, are specifically optimized for the mSUGRA model.  In order to understand how the cuts can be improved for the slepton co-NLSP scenario, it is helpful to compare the multilepton signals of mSUGRA and GMSB\@.  The details of this comparison are contained in appendix~\ref{sec:GGMvSugra}. In summary, we learn that slepton co-NLSPs produce more leptons, more missing energy, and harder leading lepton $p_T$'s.  Therefore, our main point is that when optimizing for gauge mediation one should choose cuts that are as inclusive as possible regarding the lepton multiplicity.  Backgrounds can be reduced while keeping the signal by cutting hard on missing energy and lepton $p_T$'s.

\begin{enumerate}

\item {\bf Same-sign dilepton optimization}: 
The same-sign dilepton search can be optimized for GMSB by taking advantage of the hard MET and hard lepton $p_T$ spectra of the slepton co-NLSP scenario.
We find that the search gives significantly stronger limits if a hard MET cut is added, $\met > 60$~GeV, and we used this approach when presenting the limits and reach in section~\ref{sec:results}.  Moreover, it may be possible to further optimize the search by cutting tighter on lepton $p_T$'s.

\item {\bf Trilepton optimization}:
The current CDF trilepton search loses significant acceptance for slepton co-NLSPs because of the various vetoes it includes.  For example, the fourth lepton veto and requirement that the trilepton charge not sum to $\pm 3$ removes many GMSB events with more than 3 leptons in the final state.  In addition, the $Z$ veto removes a significant fraction of signal events where pairs of leptons accidentally reconstruct the $Z$.   The cumulative effect of these vetoes degrades the trilepton search acceptance, which is the main reason that we found significantly better reach with the more inclusive approach of the same-sign dilepton search, in section~\ref{sec:results}.  The trilepton search also includes a jet veto, which we found to reduce the acceptance for slepton co-NLSPs coming from colored SUSY production (see figure~\ref{fig:gluino} and table~\ref{tab:GluinoAcceptance} above).  Our suggestion is to use a more inclusive approach, by removing the lepton, jet, and $Z$ vetoes and the charge requirement.  Instead, backgrounds can be reduced by requiring additional MET and harder first and second lepton $p_T$'s. 

\item {\bf $\ge 4$ lepton search}:
To our knowledge, the Tevatron has not searched for new physics in the $\ge 4l + \met$ channels, although such searches were suggested by some of the early GMSB literature~\cite{Dimopoulos:1996fj,Cheung:1997tg,Baer:1999tx}.  It would be very useful to conduct searches in these channels because they have low SM background and GMSB can produce many more than three leptons (as many as 6 with our benchmarks, see table~\ref{tab:topologies} and figure~\ref{fig:nlep}) in the final state.  Such a dedicated search would be particularly effective for our slepton benchmarks, where the production of $\ge 4$ leptons is common.

\item {\bf Multi-tau search}:
As we found in section~\ref{sec:future}, the same-sign dilepton and trilepton searches have poor acceptance for slepton co-NLSP scenarios with $\tau$-rich decays.   This is because these searches mostly rely on leptonically decaying $\tau$s, which requires paying a 35\% branching ratio per $\tau$.  Therefore, we recommend pursuing multiple hadronically decaying $\tau$s in the final state.  While this is more difficult experimentally, it may be possible to take advantage of the large $\met$ and large $\tau$ $p_T$'s of the slepton co-NLSP scenario.

\end{enumerate}

\subsection{Excess in SS Dileptons?}

The CDF same-sign dilepton search~\cite{D0Dilep}, described in section~\ref{sec:searches}, has looked inclusively for events in 1~fb$^{-1}$ with two leptons of the same charge.
Interestingly, the inclusive search has a slight excess, which is barely consistent within $2\sigma$.  Even more interestingly, CDF shows several distributions for events that pass the inclusive selection, including histograms of the MET and hardest lepton $p_T$.  CDF has observed four events with $\mathrm{MET}>80$~GeV when only $\sim0.9$ are expected and four events with leading lepton $p_T$ above 90 GeV where $\sim1.5$ events are expected.  It is also intriguing that Sleuth/Vista have identified like-sign dilepton final states to be among the most discrepant states, relative to background expectations, at both CDF~\cite{Aaltonen:2008vt} and D0~\cite{D0Sleuth}.

These same-sign dilepton excesses are not statistically significant.  Nevertheless, it is amusing to entertain the idea that CDF has detected a hint of supersymmetry, because, as we have seen in section \ref{sec:benchmarks}, same-sign dilepton events with high MET and a hard leading lepton are generically predicted by gauge mediation with slepton co-NLSP\@.  So in this subsection, we will attempt to answer the following question: can gauge mediation explain this slight excess at high MET and lepton $p_T$, consistently with the current experimental constraints?  (For a different possible explanation, see~\cite{delAguila:2007ua}).

In order to answer this question, we have looked at models at the edge of the current 95\% exclusion region, as determined in section~\ref{sec:results} and \ref{sec:LHC}, and counted the number of events that pass the CDF same-sign dilepton cuts with high MET or hard leading lepton $p_T$.  For simplicity, we have fixed $m_{\tilde e_R}=120$~GeV\@. The result is summarized in table \ref{tab:excess}. We find that all of our benchmark scenarios can give several events of interest.  For example, left-handed slepton production with (without) an intermediate bino gives about 4 (3) events with MET above 80 GeV\@.
We find this exciting because the trilepton searches could have already ruled out any potential explanation of the ``excess" in terms of slepton co-NLSPs. This is not the case.  We have also considered gluino production with slepton co-NLSP, which we discussed above in section~\ref{sec:LHC}.  For this channel, we find that gauge mediation could have produced as many as $\sim6$ events with high MET or a hard leading lepton, consistent with current limits.  

\begin{table}[t!]
\begin{center} \begin{tabular}{| c | c | c |}
\hline
& \multicolumn{2}{c|}{Number of Events (1 fb$^{-1}$)}\\
\cline{2-3}
channel & $\mathrm{MET}>80$~GeV&$p_T^1>90$~GeV  \\
\hline
wino &1.8  & 0.9  \\
higgsino &2.8 &2.2 \\
slepton$_L$ &3.1 &2.4  \\
slepton$_L$ +bino &3.9 &2.9  \\
gluino&5.6&6.8 \\
\hline
\end{tabular}\end{center}
\caption{The number of same-sign dilepton events with high MET or high leading lepton $p_T$, for models at the edge of the current 95\% exclusion (see section~\ref{sec:results}), fixing $m_{\tilde e_R} = 120$~GeV\@.  Such events may explain the slight excesses observed by CDF\@.
\label{tab:excess}}
\end{table}

In more detail, we have investigated the range of currently allowed production masses which could ``explain" the slight excess (by which we mean, rather arbitrarily, to have $\ge 2$ SS dilepton events with $\met>80$ GeV)\@. For higgsino production, this corresponds to production masses in the range $m_{\tilde H_1} = 190 - 210$~GeV\@.  For left handed slepton production, we find $m_{\tilde \nu} = 150 - 190$~GeV without an intermediate bino, and $m_{\tilde \nu} = 170 - 200$~GeV with an intermediate bino.  For gluino production, the mass range of interest is $m_{\tilde g} = 400 - 430$~GeV\@.  

We are encouraged to learn that all of these mass ranges are within reach at the Tevatron with 10~fb$^{-1}$ (see figures~\ref{fig:limits} and \ref{fig:gluino}).  In fact, these ranges of masses can be discovered with $5\sigma$ significance, as can be seen in figure~\ref{fig:discover}.  This means that if CDF has seen the first evidence of supersymmetry in the same-sign dilepton search, then the Tevatron still has the opportunity to make a spectacular discovery.  Alternatively, this hypothesis can be readily ruled out by upcoming data.

\section*{Acknowledgments}

 We thank  Sourabh Dube, Jiji Fan, Rob Forrest, Can Kilic, Sunil Somalwar, Scott Thomas,  Tomer Volansky, and Peter Wittich for helpful discussions. JTR is supported by an NSF graduate fellowship.  The work of DS was supported in part by DOE
grant DE-FG02-90ER40542, the William D. Loughlin membership at
the Institute for Advanced Study, and the DOE Early Career Award.

\appendix
\section{PGS Calibration}
\label{app:calibratePGS}

For event simulation in this paper, we use Pythia6.4~\cite{Sjostrand:2006za}, including showering, hadronization, and multiple interactions.  The events are then passed through PGS4~\cite{PGS} for detector simulation.  We use Suspect~\cite{Djouadi:2002ze} to calculate the superpartner spectrum and Pythia decay tables when available.  Pythia does not include the relevant two-body decays of the lightest neutralino and three-body decays of left-handed sleptons via off-shell $W$, $Z$ and neutralino.  We use a private code for these decay tables. For the left-handed slepton decays, we have implemented the three-body decay results from Refs.~\cite{Ambrosanio:1997bq,Djouadi:2000bx}. These results were originally for three-body squark decays, and three body decays of $\tilde e_R\to \tilde\tau_1+e+\tau$, but they are straightforward to adapt to all the cases of interest.

In this appendix, we describe the consistency checks and calibrations of our event simulations. We have confirmed the accuracy of our simulation of the trilepton and like-sign dilepton searches by comparing PGS output to the published acceptances for signal and some backgrounds.  We find that the dilepton search is well-simulated by PGS out-of-the-box, whereas minor modifications are necessary to reproduce the results of the trilepton search.

\subsection{SS Dileptons}

For the CDF dilepton search~\cite{Abulencia:2007rd}, described in section~\ref{sec:results},  we use PGS ``out-of-the-box" (i.e.\ without any further modifications), and we compare the acceptance of our simulation to the published acceptance for the leptonic $ZW$ and $ZZ$ backgrounds.  These are summarized in table~\ref{tab:dilepZW}. The cuts that have been applied are for the ``inclusive" analysis described in section~\ref{sec:results}. For $ZW$ where the $\met$ is real, we have also compared acceptances including an additional $\met>15$ GeV cut.  In all cases, we see that the acceptances agree with the published values to better than 25\%, indicating good agreement between our PGS simulations and the official CDF simulations. 

\begin{table}[t!]
\begin{center}
\begin{tabular}{|l|c|c|}
\hline
Process & PGS & CDF \\
\hline
$ZW$ & 0.094 & 0.080\\
$ZW$ ($\met>15$ GeV) & 0.086 & 0.070\\
$ZZ$ & 0.15 & 0.18 \\
\hline
\end{tabular}
\caption{\label{tab:dilepZW} Background $ZW$ and $ZZ$ acceptances for the CDF same-sign dilepton search~\cite{Abulencia:2007rd}.  
We compare the acceptances from our PGS simulation to the published acceptances and find good agreement.  The acceptances are normalized to leptonic ZW, $l=e,\mu,\tau$.}
\end{center}
\end{table}

\subsection{Trileptons}

It was more challenging to reproduce the CDF trilepton search~\cite{CDFTrilep3p2, Aaltonen:2008pv} using PGS\@. In the end, we found that PGS out-of-the-box was not sufficient. It is easy to understand the reason for this: since the search requires three leptons, any difference between the per-lepton PGS reconstruction efficiency, $\epsilon$, and the actual efficiency of CDF, is amplified because the acceptance scales as $\epsilon^3$.  Also, the trilepton search probes softer leptons than the dilepton search, $p_T>5$~GeV instead of  $p_T>10$~GeV\@.  Softer leptons have a lower, more $p_T$-dependent reconstruction efficiency which is not modeled correctly by PGS\@.  We find that PGS, out-of-the-box, overestimates the trilepton acceptance by about a factor of 2.

To model more accurately the CDF detector, we have modified PGS in three ways: (1) we use a crack parameterization that better matches the CDF geometry~\cite{Meade:2009qv}, (2) we have modified the lepton definitions to closer represent the ones used by the CDF trilepton search~\cite{SourabhThesis}, and (3) we have defined an isolated-track object class, as used in the CDF analysis to identify 1-prong $\tau$ decays.\footnote{The CDF analysis includes two types of electrons and two types of muons, referred to separately as {\it tight} and {\it loose} to represent more and less stringent ID requirements, respectively~\cite{SourabhThesis}.  We find that PGS leptons are not refined enough to make this differentiation.  Therefore, we use one lepton ID class, which is meant to simulate the sum of {\it tight} and {\it loose} leptons used in the CDF analysis.}

\begin{table}[t!]
\begin{center} \begin{tabular}{| c | c | c | | c  | c |}
 \hline
& \multicolumn{2}{c||}{\boldmath{$e^-e^+$}}  & \multicolumn{2}{c|}{\boldmath{$\mu^-\mu^+$}} \\
\cline{2-5}
 & $p_t \ge 20$ GeV  & $p_T < 20$ GeV & $p_t \ge 20$ GeV &  $p_T < 20$ GeV      \\
\hline 
 weight &  0.80 & 0.83 & 0.97 & 0.56 \\
 \hline
\end{tabular} \end{center}
\caption{\label{tab:weights} We define lepton weights so that PGS matches the CDF simulation in the dilepton control region.  First, the weights for $p_T \ge 20$~GeV are determined by dileptons that reconstruct the $Z$, and then the weights for $p_T < 20$~GeV are determined by the remaining Drell-Yan events, which are dominated by lepton pairs with low invariant mass.}
\end{table}

\begin{table}[t!]
\begin{center} \begin{tabular}{| c | c  | c |}
\hline
\bf Nominal Point& $3l$ & $2l+t$ \\
\hline
PGS & 8.5 & 11.8 \\
PGS (reweighted) & 4.5 & 8.1 \\
CDF & 4.6 & 6.8 \\
\hline
\end{tabular} \end{center}
\caption{The PGS, PGS with lepton weights, and CDF trilepton counts for the nominal mSUGRA point defined in \cite{Aaltonen:2008pv}.  Here, $m_0 = 60$~GeV, $m_{1/2} = 190$~GeV, $\tan \beta=3$, $A_0=0$, and $\mu>0$.    The two bins correspond to three leptons, and two leptons and one track.  \label{tab:nominal}}
\end{table}

After including these modifications, the agreement with CDF is better, but still not good enough. To understand better the per-lepton acceptances, we have examined using PGS the dilepton control sample used in the 
CDF trilepton search~\cite{SourabhThesis}. This is populated mainly by Drell-Yan production, and its acceptance should be dominated by the per-lepton acceptance.

We find that PGS overestimates the lepton acceptance relative to the CDF search.  The difference between the PGS and true efficiency depends on the lepton $p_T$, particularly for low $p_T$ muons.  Therefore, we define per-lepton weights for PGS, which we use to reweight PGS events to match the dilepton control sample. (A similar procedure is used by the CDF collaboration when tuning their detector simulator to control samples~\cite{SourabhThesis}.) We allow the lepton weights to depend on lepton $p_T$ by using two $p_T$ bins separated by $p_T=20$~GeV\@.  The lepton weights we use are shown in table~\ref{tab:weights}.

\begin{figure}[t!]
\hbox{\hspace{-0.8cm}\vbox{\includegraphics[scale=0.52]{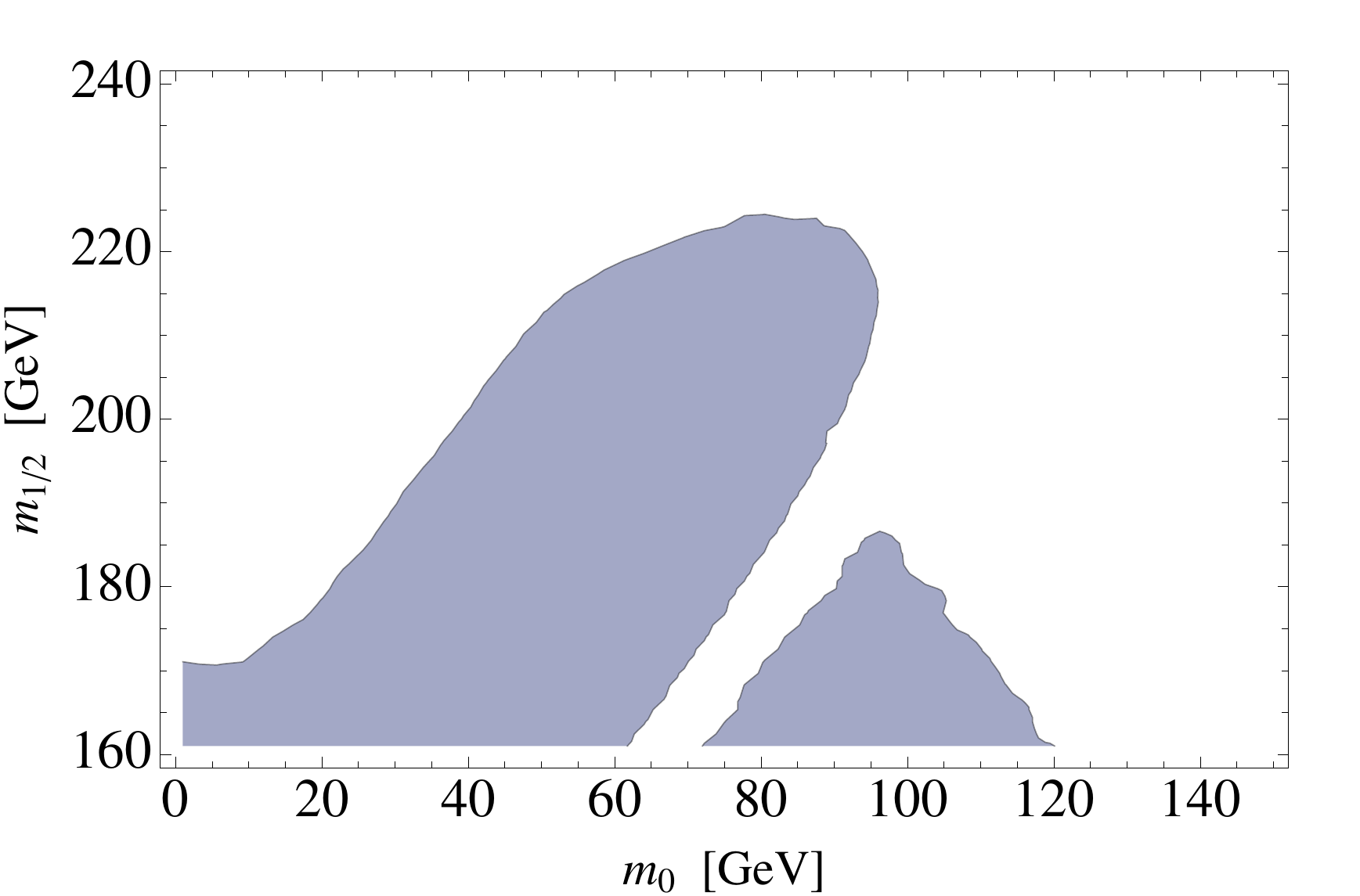}\vspace{-0.25cm} }
 \hspace{-8cm} \includegraphics[scale=0.24]{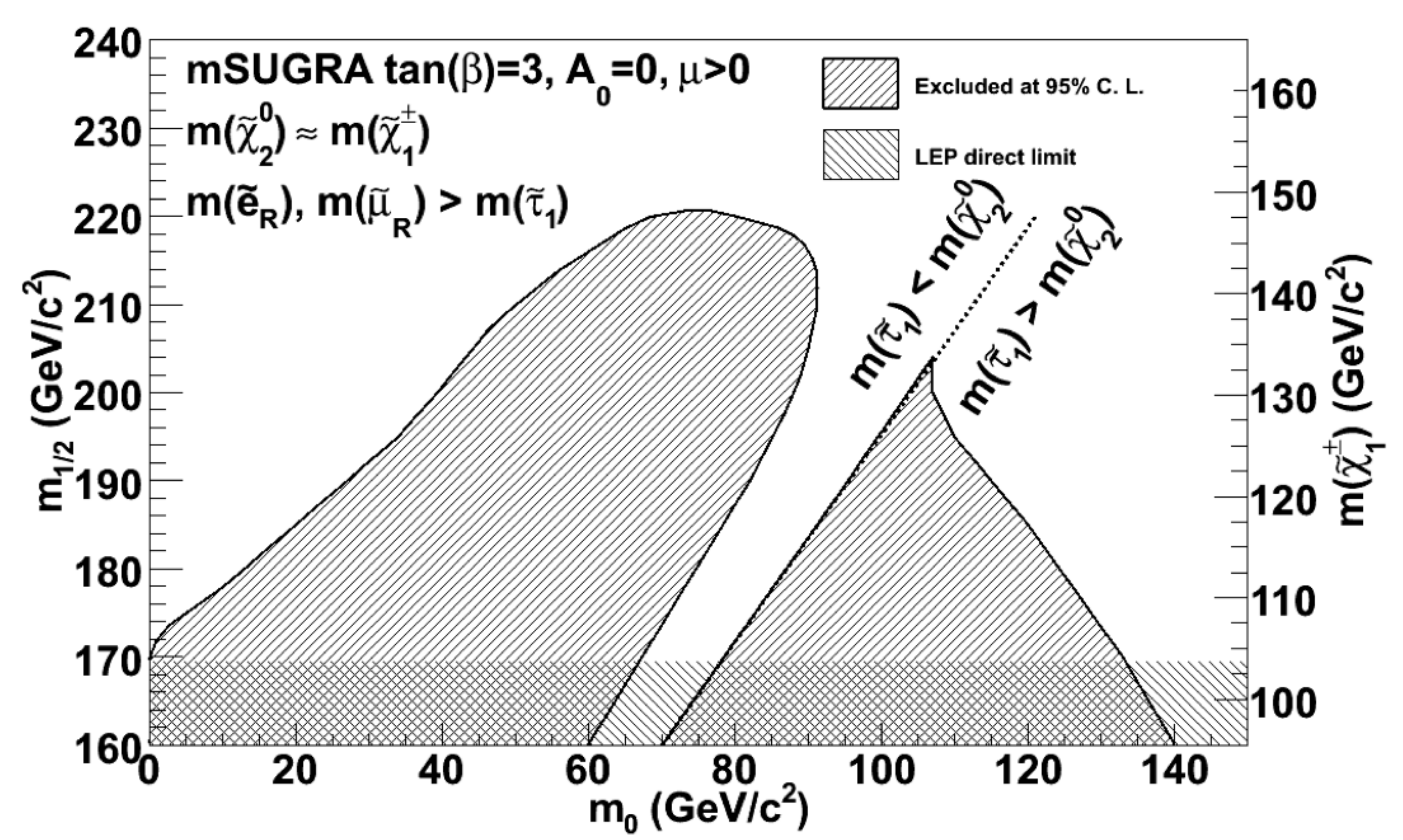}}

\caption{The {\it left} plot shows the PGS limit and the {\it right} plot shows the actual CDF limit~\cite{Aaltonen:2008pv}.  We have included the PGS lepton weights of table~\ref{tab:weights}.  We include a uniform K-factor of 1.35 to adjust the LO pythia cross-section for $\tilde N_2 + \tilde C_1$ production, to the NLO cross-section.  We have checked using Prospino \cite{Beenakker:1996ed} that the K-factor varies by less than 5\% over the parameter space. \label{fig:SugraExclusion}}
\end{figure}

With our modified and reweighted PGS in hand, we check the validity of our simulation by comparing to the published signal acceptances for the mSUGRA model studied by CDF~\cite{Aaltonen:2008pv}.  The trilepton search reports efficiencies for a ``nominal point" in mSUGRA parameter space, defined in~\cite{Aaltonen:2008pv}.  In table~\ref{tab:nominal}, we compare the PGS event count to the published value, before and after applying lepton weights.  We find that our simulation agrees to better than 20\% after applying lepton weights.  Since the weights were calibrated to the dilepton control sample, this provides a nontrivial check of our simulation.  

Finally,  in figure~\ref{fig:SugraExclusion}, we compare the 95\% limit determined with acceptances from our simulation, to the published CDF limit at 2~fb$^{-1}$, for a plane of mSUGRA parameter space.  We find that our simulation reproduces fairly well the shape of the CDF limit.  We comment that the PGS acceptance does still differ from the published acceptance by up to a factor of about 2 in regions of parameter space with at least one very soft lepton, $p_T \simeq 5$~GeV\@.

\section{Application to Minimal Gauge Mediation}
\label{sec:MGM}

So far, we have studied slepton co-NLSPs within the model-independent context of GGM\@.  It is instructive to see how Minimal Gauge Mediation (MGM)  fits within our framework, since MGM served as the basis for the original phenomenological literature on gauge mediation~\cite{Dimopoulos:1996vz,Dimopoulos:1996fj,Dimopoulos:1996yq,Cheung:1997tg,Ambrosanio:1997rv,Baer:1999tx,Qian:1998zs}.  MGM is characterized by three continuous parameters: the SUSY breaking scale $\Lambda$, the messenger mass $M$, and $\tan \beta$; and two discrete parameters: the messenger number $N$ and $\mathrm{sign}(\mu)$.  This is clearly a much smaller parameter space than that of GGM, and it greatly restricts the phenomenological possibilities. 

\begin{figure}[t!]
\begin{center} \includegraphics[scale=0.8]{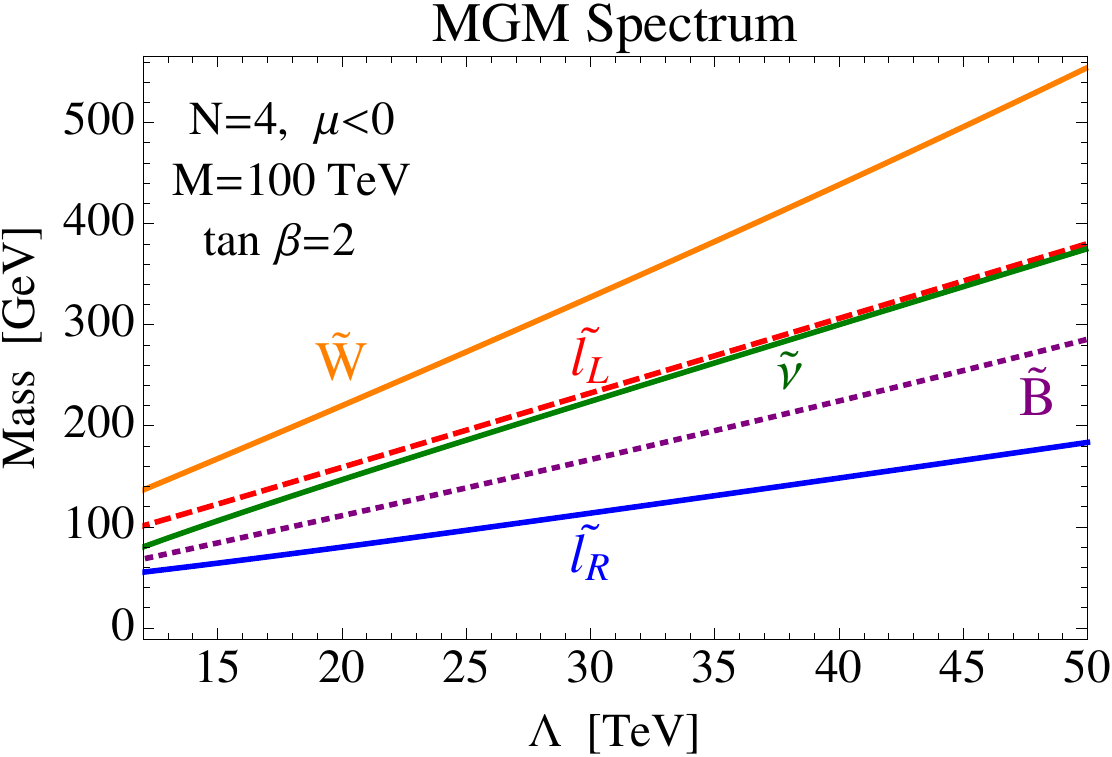} \end{center} 
\caption{A typical MGM spectrum with slepton co-NLSPs, as a function of the messenger SUSY breaking scale, $\Lambda$.  We fix the messenger number and masses, $N=4$ and $M=100$~TeV, and we choose $\tan \beta=2$ and $\mathrm{sign}(\mu)<0$.  We show the states that are relevant for the Tevatron.  Trileptons and opposite-sign dileptons result from wino production, and left-handed slepton production.  We have used SOFTSUSY~\cite{Allanach:2001kg} to calculate the spectrum.
\label{fig:MGMspec}}
\end{figure}

For instance, in MGM there are only three possible NLSPs: slepton co-NLSP, stau NLSP, or bino-like neutralino NLSP\@.\footnote{There are some special values of the MGM parameters with sneutrino NLSP (see for example the second footnote of~\cite{Ambrosanio:1997rv}).  These models are within the kinematic reach of LEP2 so are likely excluded, although this may not have been checked.}  Slepton co-NLSPs with prompt decays are generic when $N\gtrsim3$, $\tan \beta \lesssim 10$, and $M \lesssim 10^7$~GeV \cite{Dimopoulos:1996yq,Ambrosanio:1997bq}.  Larger $\tan \beta$ typically leads to stau NLSP, smaller messenger number leads to bino NLSP, and larger messenger masses imply a high SUSY breaking scale and lead to displaced decays.

In figure~\ref{fig:MGMspec}, we show a typical MGM spectrum within the prompt slepton co-NLSP regime.  The masses in MGM are proportional to $\Lambda$, shown on the $x$-axis.  Here we have fixed $N=4$, $M=100$~TeV, $\tan \beta=2$, and $\mu<0$.  We find that the spectrum is similar to our type IV benchmark of section~\ref{sec:taxonomy}, with light left-handed sleptons and an intermediate bino-like neutralino with mass between the right and left-handed sleptons.  Moreover, the lightest chargino and second lightest neutralino are both mostly wino-like, and are light enough to play a role in the Tevatron phenomenology.  The colored states in MGM are too heavy to be produced at the Tevatron, and are not shown.

\begin{figure}[t!]
\hbox{\hspace{-0.7cm} \includegraphics[scale=0.35]{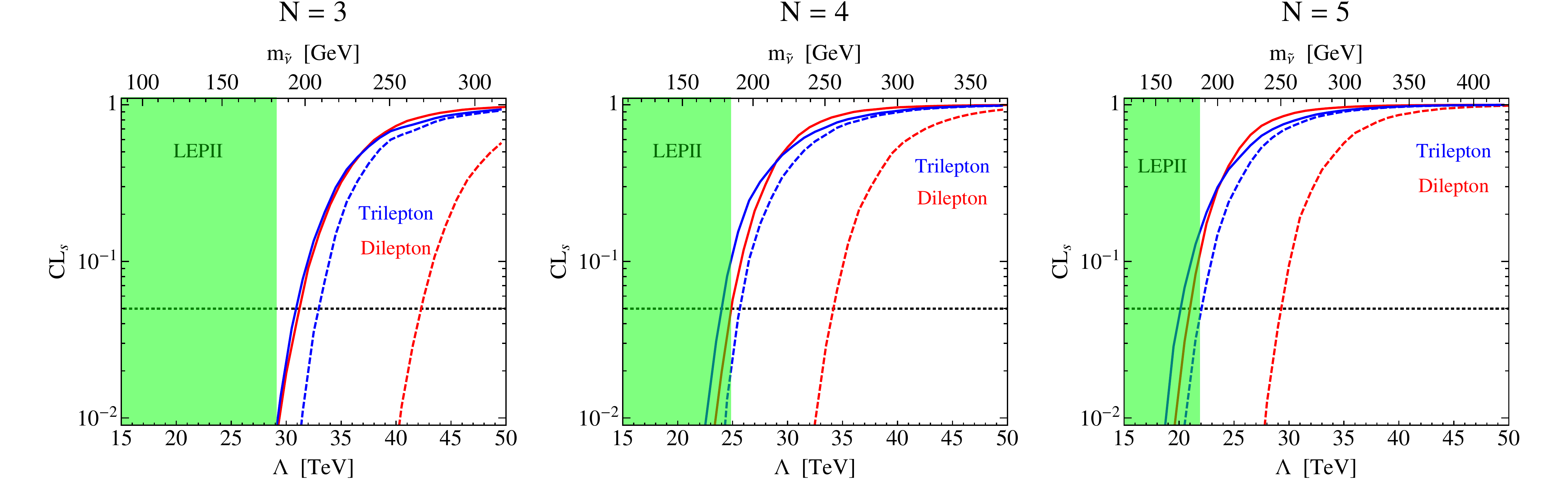}  }
\caption{The CL$_s$ statistics for the CDF trilepton search (blue) and SS dilepton search (red) are shown for MGM lines with messenger numbers $N=3,4,5$. The solid (dashed) lines indicate current (projected 10 fb$^{-1}$) searches. The dotted black lines indicate the 95\% confidence level. Included also in the plots are the LEP II limits on direct right-handed slepton production, which are stronger than the current Tevatron limits except for $N=3$.
\label{fig:MGMlimit}}
\end{figure}

MGM with slepton co-NLSP can populate like-sign dilepton and trilepton final states when left-handed sleptons are produced, as in the type IV benchmark, or when winos are produced.  Winos will decay to the left-handed sleptons, resulting in final states that are similar to left-handed slepton production, augmented by additional leptons from the wino decay.  We show the current and projected (10~fb$^{-1}$) limits on MGM in figure~\ref{fig:MGMlimit}, fixing $M=100$~TeV, $\tan \beta=2$, $\mu < 0$, and varying $\Lambda$ and $N$.  For $N=5$, the current Tevatron limit is worse than the LEP II limit on direct $\tilde \mu_R \tilde \mu_R^*$ production, of $m_{\tilde l_R}>96$~GeV\@.  This follows from the fixed mass ratios between $\tilde l_R$ and the heavier states, a restriction that is not present in GGM\@.
For $N=4$, LEP and Tevatron are competitive, and for $N=3$, Tevatron sets a slightly better limit than LEP\@.  Tevatron gains a competitive advantage at smaller $N$ because the winos are lighter, relative to the right-handed sleptons, leading to a larger cross-section at the Tevatron.  For all three values of $N$, we find that the projected reach at Tevatron will surpass the LEP II limit.

\begin{figure}[t!]
\begin{center} \includegraphics[scale=0.52]{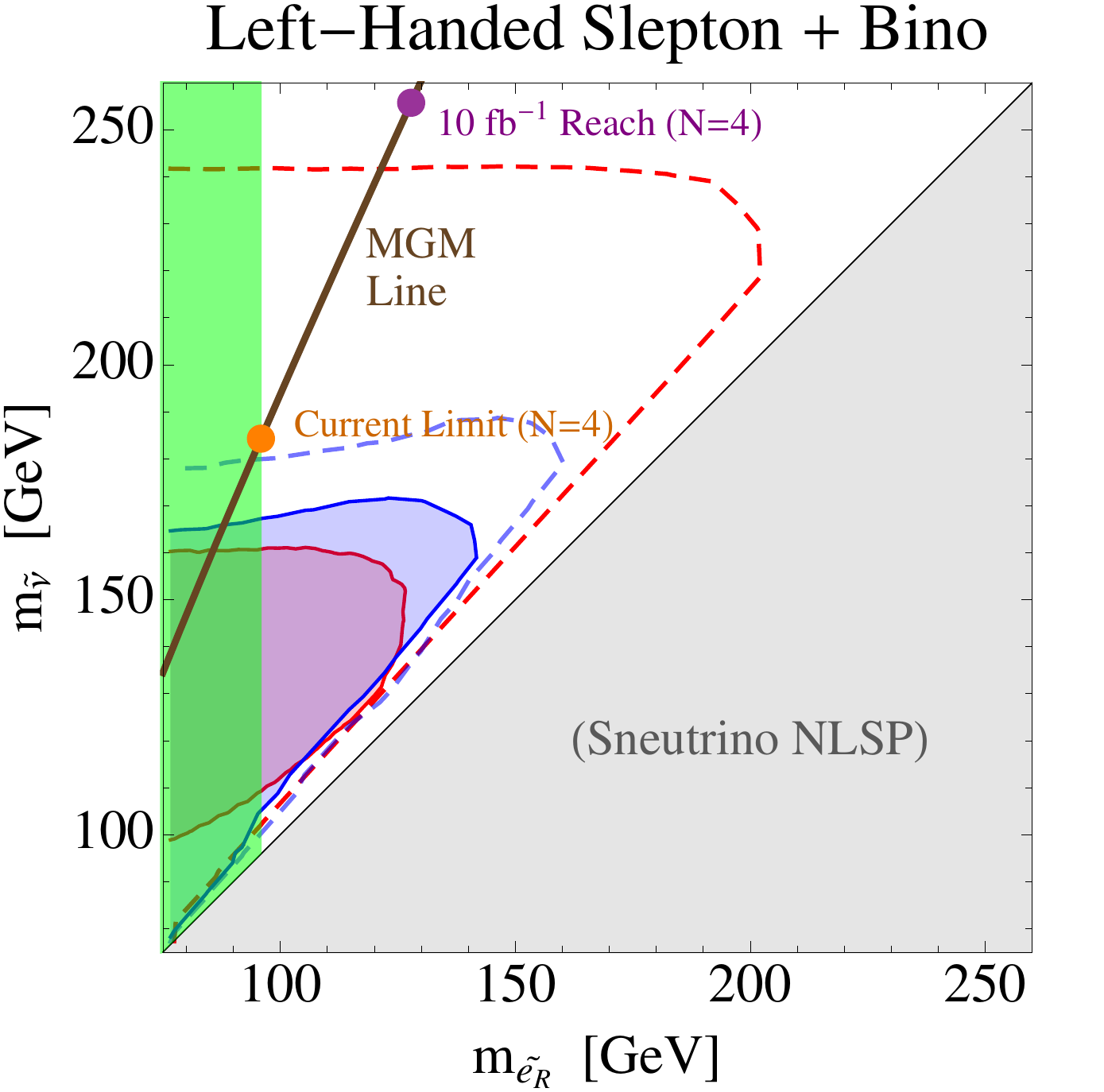} \end{center} 
\caption{The MGM mass line superimposed on the exclusion and reach contours for left-handed slepton production with an intermediate bino, reproduced from figure~\ref{fig:limits}.  MGM imposes a relation between the left-handed and right-handed slepton masses, corresponding to the brown line for $\tan \beta = 2$.    MGM also includes a light wino, and wino production gives additional events. This leads to a slightly stronger limit (orange circle) and reach (purple circle), as shown here for messenger number $N=4$.} 
\label{fig:MGMon2d}
\end{figure}

It is possible to set a conservative limit on MGM by considering left-handed slepton production alone.  For this purpose, we can directly apply our results from section~\ref{sec:results}.   We illustrate this procedure in figure~\ref{fig:MGMon2d}, where we show the MGM mass line, superimposed on the type IV limit plot of section~\ref{sec:results}.  This mass line, $m_{\tilde l_L} \simeq 2 \, m_{\tilde l_R}$, applies robustly to MGM with $\tan \beta =2$, regardless of messenger number or messenger mass in the prompt regime, $M \lesssim 10^7$~GeV\@.  The limit and reach on left-handed slepton production correspond to when the MGM mass line crosses the relevant limit and reach curves on this plot.  The actual limit and reach, including wino production is about 10\% stronger in mass, and shown on the plot.

\section{GGM vs.\ mSUGRA}
\label{sec:GGMvSugra}

We include a brief comparison of the multilepton signals from GMSB with slepton co-NLSP to those from mSUGRA (for a review with references see~\cite{Martin:1997ns}).  This comparison is useful because many Tevatron searches for SUSY, such as the trilepton search described above, were designed with the mSUGRA signal in mind.  This has led the searches to be biased in various ways. 

\begin{figure}[!t]
\begin{center}
\includegraphics[scale=0.5]{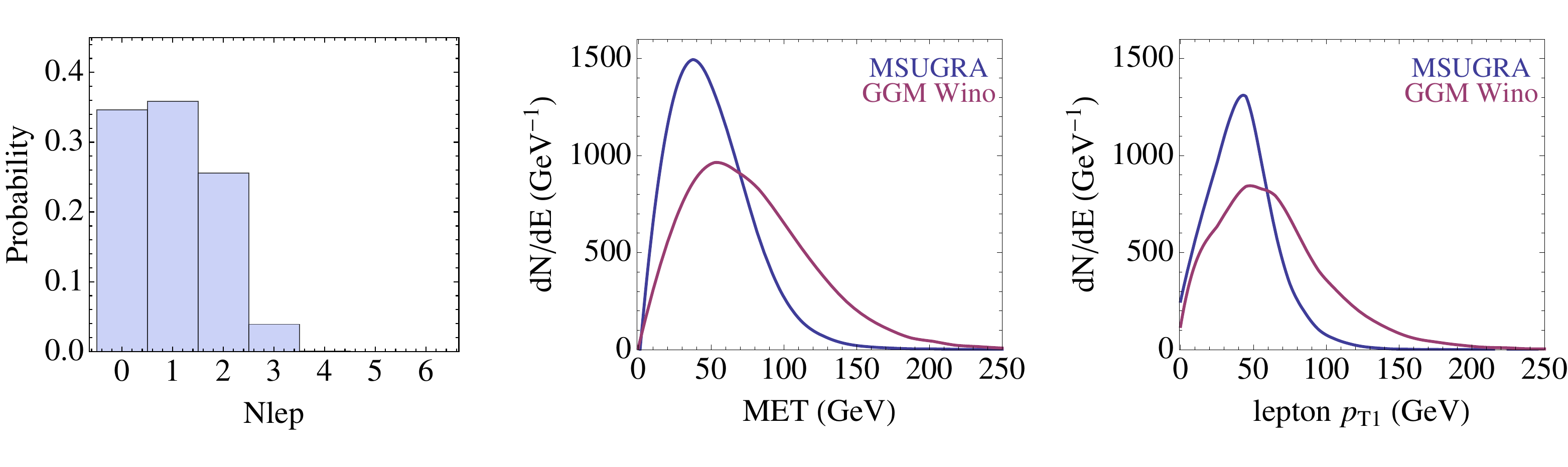} 
\caption{mSUGRA comparison.  For mSUGRA we choose a nominal point, $\tan\beta=3$, $\mu>0$, $A=0$, $m_0=100$ GeV and $m_{1/2}=265$ GeV\@.  On the left we show the number of leptons with $|\eta|<1$ and $p_T>10$ GeV\@.  Comparing with figure~\ref{fig:nlep}, we see that GMSB tends to produce more leptons than mSUGRA\@.  The center and right plots compare the distributions of missing energy and hardest lepton $p_T$ (with 100k events) between mSUGRA (blue) and GGM with wino production (red).  For GGM we match the wino and right-handed slepton masses with mSUGRA, $m_{\tilde W}\simeq 200$~GeV and $m_{\tilde e_R}\simeq 150$~GeV\@.}
\label{fig:msugracomp}
\end{center}
\end{figure}

Shown in fig.\ \ref{fig:msugracomp} are the number of leptons, the $\met$, and the hardest lepton $p_T$ distributions, for a nominal mSUGRA point with $\tan\beta=3$, $\mu>0$, $A=0$, $m_0=100$ GeV and $m_{1/2}=265$ GeV\@. This corresponds to a wino mass of $\sim 200$ GeV, a bino mass of $\sim 100$~GeV, and a right-handed slepton mass of $\sim 150$ GeV\@. The left-handed sleptons are heavier than the wino and are effectively decoupled. The only production mode for multiple leptons here is $\tilde\chi_1^\pm \tilde\chi_2$. These states decay with nearly 100\% branching fraction (and nearly flavor democratically) to the right-handed sleptons, which then decay to the bino LSP\@. So this scenario is most analogous to our wino production scenario, with the only difference being that the LSP is 100~GeV here instead of the massless gravitino.

In fig.\ \ref{fig:msugracomp}, we have also included for comparison the $\met$ and $p_{T1}$ distributions for the wino production GMSB scenario with the same wino and right-handed slepton mass. We see that the lepton distribution is quite similar to the that of the wino production scenario (the left-most plot in fig.\ \ref{fig:nlep}).  Of the slepton co-NLSP scenarios we have considered, wino production leads to the fewest number of leptons, which means that GMSB can produce significantly more leptons in the final state than mSUGRA\@.  We can also see from figure~\ref{fig:msugracomp} that the $\met$ and $p_{T1}$ distributions for the GMSB scenario are relatively harder, since the LSP is massless.  This suggests that when optimizing for GMSB signatures, as opposed to mSUGRA signatures, one should prepare for the possibility of more leptons in the final state and one can cut harder on $\met$ and lepton $p_T$, perhaps relaxing other cuts or vetoes as a consequence.  For our suggestions in this direction, see section~\ref{sec:discussion}.



\begin{thebibliography}{99}
  
\bibitem{Martin:1997ns}
  S.~P.~Martin,
  ``A Supersymmetry Primer,''
  arXiv:hep-ph/9709356.
  
\bibitem{Giudice:1998bp}
  G.~F.~Giudice and R.~Rattazzi,
  ``Theories with gauge-mediated supersymmetry breaking,''
  Phys.\ Rept.\  {\bf 322}, 419 (1999)
  [arXiv:hep-ph/9801271].

\bibitem{Dimopoulos:1996vz}
  S.~Dimopoulos, M.~Dine, S.~Raby and S.~D.~Thomas,
  ``Experimental Signatures of Low Energy Gauge Mediated Supersymmetry
  Breaking,''
  Phys.\ Rev.\ Lett.\  {\bf 76}, 3494 (1996)
  [arXiv:hep-ph/9601367].

\bibitem{Dimopoulos:1996fj}
  S.~Dimopoulos, M.~Dine, S.~Raby, S.~D.~Thomas and J.~D.~Wells,
  ``Phenomenological implications of low energy supersymmetry breaking,''
  Nucl.\ Phys.\ Proc.\ Suppl.\  {\bf 52A}, 38 (1997)
  [arXiv:hep-ph/9607450].

\bibitem{Dimopoulos:1996yq}
  S.~Dimopoulos, S.~D.~Thomas and J.~D.~Wells,
  ``Sparticle spectroscopy and electroweak symmetry breaking with
  gauge-mediated supersymmetry breaking,''
  Nucl.\ Phys.\  B {\bf 488}, 39 (1997)
  [arXiv:hep-ph/9609434].
  
\bibitem{Ambrosanio:1997rv}
  S.~Ambrosanio, G.~D.~Kribs and S.~P.~Martin,
  ``Signals for gauge-mediated supersymmetry breaking models at the CERN  LEP2
  collider,''
  Phys.\ Rev.\  D {\bf 56}, 1761 (1997)
  [arXiv:hep-ph/9703211].
  
\bibitem{Cheung:1997tg}
  K.~Cheung, D.~A.~Dicus, B.~Dutta and S.~Nandi,
  ``Multilepton signatures of gauge mediated SUSY breaking at LEPII,''
  Phys.\ Rev.\  D {\bf 58}, 015008 (1998)
  [arXiv:hep-ph/9711216].
  
\bibitem{Baer:1999tx}
  H.~Baer, P.~G.~Mercadante, X.~Tata and Y.~l.~Wang,
  ``The Reach of Tevatron upgrades in gauge mediated supersymmetry breaking
  models,''
  Phys.\ Rev.\  D {\bf 60}, 055001 (1999)
  [arXiv:hep-ph/9903333].
  
\bibitem{Qian:1998zs}
  J.~Qian  [D0 Collaboration],
  ``Sensitivity to gauge-mediated supersymmetry breaking models of the
  Fermilab upgraded Tevatron collider,''
  arXiv:hep-ph/9903548.


\bibitem{Culbertson:2000am}
  R.~L.~Culbertson {\it et al.}  [SUSY Working Group Collaboration],
  ``Low scale and gauge mediated supersymmetry breaking at the Fermilab
  Tevatron Run II,''
  arXiv:hep-ph/0008070.

\bibitem{Meade:2008wd}
  P.~Meade, N.~Seiberg and D.~Shih,
  ``General Gauge Mediation,''
  Prog.\ Theor.\ Phys.\ Suppl.\  {\bf 177}, 143 (2009)
  [arXiv:0801.3278 [hep-ph]].
  
\bibitem{Buican:2008ws}
  M.~Buican, P.~Meade, N.~Seiberg and D.~Shih,
  ``Exploring General Gauge Mediation,''
  JHEP {\bf 0903}, 016 (2009)
  [arXiv:0812.3668 [hep-ph]].
  
  \bibitem{D0Dilep}
D0 Collaboration, 
``Search for the associated production of charginos and neutralinos in the like sign dimuon channel,"
{\it Public Note} {\bf 5126},  (2006).

 
\bibitem{Abulencia:2007rd}
  A.~Abulencia {\it et al.}  [CDF Collaboration],
  ``Inclusive search for new physics with like-sign dilepton events in $p
  \bar{p}$ collisions at $\sqrt{s}$ = 1.96-TeV,''
  Phys.\ Rev.\ Lett.\  {\bf 98}, 221803 (2007)
  [arXiv:hep-ex/0702051].

 \bibitem{CDFTrilep3p2}
CDF Collaboration, 
``Update of the Unified Trilepton Search with 3.2 fb-1 of Data",
 {\it Public Note} {\bf 9817}, (2009).

\bibitem{Aaltonen:2008pv}
  T.~Aaltonen {\it et al.}  [CDF Collaboration],
  ``Search for Supersymmetry in $p \bar{p}$ Collisions at $\sqrt{s}$ = 1.96-TeV
  Using the Trilepton Signature of Chargino-Neutralino Production,''
  Phys.\ Rev.\ Lett.\  {\bf 101}, 251801 (2008)
  [arXiv:0808.2446 [hep-ex]].


  

\bibitem{Abazov:2009zi}
  V.~M.~Abazov {\it et al.}  [D0 Collaboration],
  ``Search for associated production of charginos and neutralinos in the
  trilepton final state using 2.3 fb-1 of data,''
  Phys.\ Lett.\  B {\bf 680}, 34 (2009)
  [arXiv:0901.0646 [hep-ex]].



\bibitem{Campbell:2010ff}
  J.~M.~Campbell and R.~K.~Ellis,
  ``MCFM for the Tevatron and the LHC,''
  arXiv:1007.3492 [hep-ph].

\bibitem{Meade:2009qv}
  P.~Meade, M.~Reece and D.~Shih,
  ``Prompt Decays of General Neutralino NLSPs at the Tevatron,''
  arXiv:0911.4130 [hep-ph].

\bibitem{Meade:2010ji}
  P.~Meade, M.~Reece and D.~Shih,
  ``Long-Lived Neutralino NLSPs,''
  arXiv:1006.4575 [hep-ph].
  
\bibitem{Carpenter:2008he}
  L.~M.~Carpenter,
  ``Surveying the Phenomenology of General Gauge Mediation,''
  arXiv:0812.2051 [hep-ph].
  
\bibitem{Rajaraman:2009ga}
  A.~Rajaraman, Y.~Shirman, J.~Smidt and F.~Yu,
  ``Parameter Space of General Gauge Mediation,''
  Phys.\ Lett.\  B {\bf 678}, 367 (2009)
  [arXiv:0903.0668 [hep-ph]].
  
\bibitem{Abel:2009ve}
  S.~Abel, M.~J.~Dolan, J.~Jaeckel and V.~V.~Khoze,
  ``Phenomenology of Pure General Gauge Mediation,''
  JHEP {\bf 0912}, 001 (2009)
  [arXiv:0910.2674 [hep-ph]].
  
\bibitem{Katz:2009qx}
  A.~Katz and B.~Tweedie,
  ``Signals of a Sneutrino (N)LSP at the LHC,''
  Phys.\ Rev.\  D {\bf 81}, 035012 (2010)
  [arXiv:0911.4132 [hep-ph]].
  
\bibitem{Katz:2010xg}
  A.~Katz and B.~Tweedie,
  ``Leptophilic Signals of a Sneutrino (N)LSP and Flavor Biases from
  Flavor-Blind SUSY,''
  Phys.\ Rev.\  D {\bf 81}, 115003 (2010)
  [arXiv:1003.5664 [hep-ph]].
  
\bibitem{Abel:2010vb}
  S.~Abel, M.~J.~Dolan, J.~Jaeckel and V.~V.~Khoze,
  ``Pure General Gauge Mediation for Early LHC Searches,''
  arXiv:1009.1164 [hep-ph].

\bibitem{Sjostrand:2006za}
  T.~Sjostrand, S.~Mrenna and P.~Z.~Skands,
  ``PYTHIA 6.4 Physics and Manual,''
  JHEP {\bf 0605}, 026 (2006)
  [arXiv:hep-ph/0603175].

\bibitem{PGS}
J. Conway, PGS: Pretty Good Simulator, \\ http://www.physics.ucdavis.edu/conway/research/software/pgs/pgs4-general.htm.


\bibitem{Djouadi:2002ze}
  A.~Djouadi, J.~L.~Kneur and G.~Moultaka,
  ``SuSpect: A Fortran code for the supersymmetric and Higgs particle spectrum
  in the MSSM,''
  Comput.\ Phys.\ Commun.\  {\bf 176}, 426 (2007)
  [arXiv:hep-ph/0211331].
  


  \bibitem{JoshDavidLHC}
  J.T.~Ruderman and D.~Shih, ``Slepton co-NLSPs at the LHC," {\it to appear}.

\bibitem{Komargodski:2008ax}
  Z.~Komargodski and N.~Seiberg,
  ``mu and General Gauge Mediation,''
  JHEP {\bf 0903}, 072 (2009)
  [arXiv:0812.3900 [hep-ph]].

\bibitem{Ambrosanio:1997bq}
  S.~Ambrosanio, G.~D.~Kribs and S.~P.~Martin,
  ``Three-body decays of selectrons and smuons in low-energy supersymmetry
  breaking models,''
  Nucl.\ Phys.\  B {\bf 516}, 55 (1998)
  [arXiv:hep-ph/9710217].
  

 \bibitem{LEP2}
 LEPSUSYWG, ALEPH, DELPHI, L3 and OPAL experiments, note LEPSUSYWG/02-09.2, http://lepsusy.web.cern.ch/lepsusy/Welcome.html

\bibitem{Dube:2008kf}
  S.~Dube, J.~Glatzer, S.~Somalwar, A.~Sood and S.~Thomas,
  ``Addressing the Multi-Channel Inverse Problem at High Energy Colliders: A
  Model Independent Approach to the Search for New Physics with Trileptons,''
  arXiv:0808.1605 [hep-ph].


\bibitem{Aaltonen:2009nr}
  T.~Aaltonen {\it et al.}  [CDF Collaboration],
  ``Search for New Bottomlike Quark Pair Decays $Q\bar Q \to
  (t W^\mp)(\bar t W^\pm)$ in Same-Charge Dilepton Events,''
  Phys.\ Rev.\ Lett.\  {\bf 104}, 091801 (2010)
  [arXiv:0912.1057 [hep-ex]].

 \bibitem{SourabhThesis}
S. S. Dube, ÒSearch for supersymmetry at the Tevatron using the trilepton signature,Ó FERMILAB-THESIS-
2008-45.

  
\bibitem{Junk:1999kv}
  T.~Junk,
  ``Confidence Level Computation for Combining Searches with Small
  Statistics,''
  Nucl.\ Instrum.\ Meth.\  A {\bf 434}, 435 (1999)
  [arXiv:hep-ex/9902006].

\bibitem{Baer:2000pe}
  H.~Baer, P.~G.~Mercadante, X.~Tata and Y.~l.~Wang,
  ``The Reach of the CERN large hadron collider for gauge mediated
  supersymmetry breaking models,''
  Phys.\ Rev.\  D {\bf 62}, 095007 (2000)
  [arXiv:hep-ph/0004001].

\bibitem{DeSimone:2008gm}
  A.~De Simone, J.~Fan, M.~Schmaltz and W.~Skiba,
  ``Low-scale gaugino mediation, lots of leptons at the LHC,''
  Phys.\ Rev.\  D {\bf 78}, 095010 (2008)
  [arXiv:0808.2052 [hep-ph]].

\bibitem{DeSimone:2009ws}
  A.~De Simone, J.~Fan, V.~Sanz and W.~Skiba,
  ``Leptogenic Supersymmetry,''
  Phys.\ Rev.\  D {\bf 80}, 035010 (2009)
  [arXiv:0903.5305 [hep-ph]].

\bibitem{Beenakker:1996ed}
  W.~Beenakker, R.~Hopker and M.~Spira,
  ``PROSPINO: A program for the PROduction of Supersymmetric Particles In
  Next-to-leading Order QCD,''
  arXiv:hep-ph/9611232.
  
\bibitem{Beenakker:1996ch}
  W.~Beenakker, R.~Hopker, M.~Spira and P.~M.~Zerwas,
  ``Squark and gluino production at hadron colliders,''
  Nucl.\ Phys.\  B {\bf 492}, 51 (1997)
  [arXiv:hep-ph/9610490].

\bibitem{Aaltonen:2008vt}
  T.~Aaltonen {\it et al.}  [CDF Collaboration],
  ``Global Search for New Physics with 2.0/fb at CDF,''
  Phys.\ Rev.\  D {\bf 79}, 011101 (2009)
  [arXiv:0809.3781 [hep-ex]].
  
 \bibitem{D0Sleuth}
D0 Collaboration,
``Model Independent Search for New Physics at D0 in Final States Containing Leptons,"
 {\it Public Note} {\bf 5777},  (2009).

\bibitem{delAguila:2007ua}
  F.~del Aguila and J.~A.~Aguilar-Saavedra,
  ``Like-sign dilepton signals from a leptophobic $Z^\prime$ boson,''
  JHEP {\bf 0711}, 072 (2007)
  [arXiv:0705.4117 [hep-ph]].

\bibitem{Djouadi:2000bx}
  A.~Djouadi and Y.~Mambrini,
  ``Three-body decays of top and bottom squarks,''
  Phys.\ Rev.\  D {\bf 63}, 115005 (2001)
  [arXiv:hep-ph/0011364].


\bibitem{Allanach:2001kg}
  B.~C.~Allanach,
  ``SOFTSUSY: a program for calculating supersymmetric spectra,''
  Comput.\ Phys.\ Commun.\  {\bf 143}, 305 (2002)
  [arXiv:hep-ph/0104145].
  
 
  
 \end{thebibliography}
\end{document}